\documentclass[letterpaper]{article}
\pdfoutput=1
\usepackage{aaai}
\usepackage{times}
\usepackage{helvet}
\usepackage{courier}
\usepackage{savesym}
\usepackage{multirow}
\usepackage{url}
\usepackage[titletoc]{appendix}
\usepackage{xspace}
\usepackage[tight]{subfigure}
\usepackage{graphicx}
\usepackage{url}
\usepackage{hyperref}
\usepackage{enumitem}
\usepackage{color}
\makeatletter
\newif\if@restonecol
\makeatother

\usepackage[lined, boxed, ruled, commentsnumbered, noend]{algorithm2e}
\usepackage{amsmath,amssymb,xfrac}
\usepackage{amsthm}
\setitemize{noitemsep,topsep=1pt,parsep=1pt,partopsep=1pt, leftmargin=12pt}
\newtheorem{lemma}{Lemma}
\newtheorem{theorem}{Theorem}
\newtheorem{proposition}{Proposition}

\makeatletter

\newcommand{\Rmnum}[1]{\expandafter\@slowromancap\romannumeral #1@}
\makeatother
\usepackage{caption}

\newcommand{\ia}{a}
\newcommand{\icA}{{\mathcal{A}}}
\newcommand{\iv}{v}
\newcommand{\icV}{{\mathcal{V}}}
\newcommand{\id}{d}

\newcommand{\iobjf}{f}

\newcommand{\iw}{w}
\newcommand{\icW}{{\mathcal{W}}}
\newcommand{\icO}{{\mathcal{O}}}
\newcommand{\is}{s}
\newcommand{\icS}{{\mathcal{S}}}
\newcommand{\iobjg}{g}
\newcommand{\ifmax}{f_{\max}}
\newcommand{\iobs}{y}
\newcommand{\ibobs}{{\mathbf{y}}}

\newcommand{\iObs}{Y}
\newcommand{\ibObs}{{\mathbf{Y}}}

\newcommand{\iprobObs}{P}

\newcommand{\iExp}{\mathbb{E}}

\newcommand{\icost}{c}

\newcommand{\ibid}{b}
\newcommand{\ibidalt}{b'}
\newcommand{\iBid}{B}
\newcommand{\iBidalt}{B'}
\newcommand{\icmax}{c_{\max}}
\newcommand{\icmin}{c_{\min}}
\newcommand{\iReal}{{\mathbb{R}}}

\newcommand{\icB}{{\mathcal{B}}}
\newcommand{\iMechanism}{{\mathcal{M}}}

\newcommand{\ipolicy}{\pi}
\newcommand{\iPolicy}{\boldsymbol{\pi}}
\newcommand{\ipayment}{\theta}
\newcommand{\iPayment}{\boldsymbol{\theta}}
\newcommand{\ipaymentconst}{\theta^d}

\newcommand{\ipaymentconstr}{\theta^{d,r}}

\newcommand{\iP}{P}
\newcommand{\icSalt}{{\mathcal{S'}}}
\newcommand{\icWalt}{{\mathcal{W'}}}
\newcommand{\iwstar}{w^*}
\newcommand{\icT}{{\mathcal{T}}}

\newcommand{\imarginal}{\Delta}
\newcommand{\iMarginal}{\boldsymbol{\Delta}}
\newcommand{\imarginalalt}{\Delta'}
\newcommand{\iMarginalalt}{\boldsymbol{\Delta'}}
\newcommand{\irho}{\rho}
\newcommand{\ialpha}{\alpha}

\newcommand{\ii}{i}

\newcommand{\iij}{j}

\newcommand{\ik}{k}
\newcommand{\ikalt}{k'}
\newcommand{\isalt}{s'}
\newcommand{\ibobsSet}{{\mathbf{Z}}}
\newcommand{\ir}{r}
\newcommand{\iZ}{Z}
\newcommand{\iin}{n}

\newcommand{\icTalt}{{\mathcal{T'}}}
\newcommand{\icR}{{\mathcal{R}}}

\newcommand{\il}{l}
\newcommand{\iy}{y}
\newcommand{\ix}{x}
\newcommand{\iX}{X}
\newcommand{\isgreedypol}{\ipolicy_{G}}
\newcommand{\istgreedypol}{\ipolicy_{TG}}
\newcommand{\isoptpol}{\ipolicy_{OPT}}
\newcommand{\iobjgavg}{g_{avg}}

\newcommand{\greedy}{\textsc{Greedy}\xspace}
\newcommand{\tgreedy}{\textsc{TGreedy}\xspace}
\newcommand{\opt}{\textsc{Opt}\xspace}
\newcommand{\cgreedy}{\textsc{ConstGreedy}\xspace}
\newcommand{\ctgreedy}{\textsc{ConstTGreedy}\xspace}

\newcommand{\sgreedy}{\textsc{SeqGreedy}\xspace}
\newcommand{\stgreedy}{\textsc{SeqTGreedy}\xspace}
\newcommand{\sopt}{\textsc{SeqOpt}\xspace}
\newcommand{\random}{\textsc{Random}\xspace}

\newcommand{\citet}[1]{\citeauthor{#1} (\citeyear{#1})}

\title{Incentives for Privacy Tradeoff in Community Sensing}
\author{Adish Singla \and Andreas Krause\\
ETH Zurich\\
Universit\"atstrasse 6, 8092 Z\"{u}rich, Switzerland\\
}
\begin{document}
%

\maketitle
\begin{abstract}
\begin{quote}
Community sensing, fusing information from populations of privately-held sensors, presents a great opportunity to create efficient and cost-effective sensing applications. Yet, reasonable privacy concerns often limit the access to such data streams. 
How should systems valuate and negotiate access to private information, for example in return for monetary incentives?
How should they optimally choose the participants from a large population of strategic users with privacy concerns, and compensate them for information shared?

In this paper, we address these questions and present a novel mechanism, \stgreedy, 
for budgeted recruitment of participants in community sensing. 
We first show that privacy tradeoffs in community sensing can be cast as an adaptive submodular optimization problem. We then 
design a budget feasible, incentive compatible (truthful) mechanism for adaptive submodular maximization, which achieves near-optimal utility for a large class of sensing applications. 
This mechanism is general, and of independent interest.
%
We demonstrate the effectiveness 
of our approach in a case study of air quality monitoring, 
using data collected from the Mechanical Turk platform. Compared to the state of the art, our approach achieves up to 
30\% reduction in cost in order to achieve a desired level of utility.
\end{quote}
\end{abstract}
\section{Introduction}\label{sec:introduction}
Community sensing is a new paradigm for creating efficient and cost-effective sensing applications by harnessing the data of large populations of sensors. 
For example, the accelerometer data from smartphone users could be used for earthquake detection and fine grained analysis of seismic events. 
Velocity data from GPS devices (in smartphones or automobiles) could be used to provide real-time traffic maps or detect accidents. However, accessing this stream of private sensor data raises reasonable concerns about privacy of the individual users. For example, mobility patterns and the house or office locations of a user could possibly be inferred from their GPS tracks \cite{2007-pervasive_inference-attack}. Beyond concerns about sharing sensitive information, there are general anxieties among users about sharing data from their private devices. 
These concerns limit the practical applicability of deploying such applications. In this paper, we propose a principled approach to negotiate access to certain private information in an incentive-compatible manner.

{\bf Applications of community sensing } \looseness -1 are numerous.  
Several case studies have demonstrated the principal feasibility and usefulness of community sensing. 
A number of research and commercial prototypes are build, often relying on special campaigns to recruit volunteers \cite{2010-ieee_msr_geolife} or on contracts with service providers to obtain anonymized data \cite{2007-techreport_travel-time-estimation}. The SenseWeb system \cite{2007-ieee_senseweb} has been developed as an infrastructure for sharing sensing data to enable various applications. Methods have been developed to estimate traffic \cite{2007-mobisys_street-traffic-estimation,2008-online_mobile-millennium,2008-ispn_krause_community-sensing},
 perform forecasts about future traffic situations \cite{2005-uai_horvitz_traffic} or predict a driver's trajectory \cite{2006-ubicomp_predestination}. Cell tower signals obtained from the service providers are leveraged for travel time estimation on roadways \cite{2007-techreport_travel-time-estimation}. 
Additionally, captured images and video clips from smartphones have been used to link places with various categories \cite{2012-ubicomp_cps}. 
\citet{2012-geophysics_krause_community-seismic-network} describes the design of a \emph{Community Seismic Network} to detect and monitor earthquakes using a dense network of low cost sensors hosted by volunteers from the community. \citet{2010-iwgs_opensense} envisions a community driven sensing infrastructure for monitoring air quality.
%

{\bf Privacy concerns in community sensing }
are expected and reasonable \cite{2007-_modot,2007-techreport_travel-time-estimation,2005-chi_privacy-preferences}. Irrespective of the  models of privacy we consider \cite{2002-journal-ufks_k-anonymity,2006-icalp_differential-privacy,2006-icde_l-diversity}, the key concern is about identifiability as users become members of increasingly smaller groups of people sharing the same characteristics inferred from data. Beyond general anxieties about the sharing of location and mobility data, studies have demonstrated that, even with significant attempts at obfuscation, home and work locations of drivers can be inferred from GPS tracks \cite{2007-pervasive_inference-attack}. 

{\bf Incentives to participants for privacy tradeoff. } 
\citet{2005-chi_privacy-preferences} show that people's willingness to share information depends greatly on the type of information being shared, with whom the information is shared, and how it is going to be used. They are willing to share certain private information if compensated in terms of their utility gain \cite{2008-aaai_krause_privacy-personalization}. 
In this paper, we are exploring the design of intelligent systems that empower users to consciously share certain private information in return of, \emph{e.g.}, monetary or other form of incentives. We model the users as strategic agents who are willing to negotiate access to certain private information, aiming to maximize the monetary incentives they receive in return. Empowering users to opt into such negotiations is the key idea that we explore in this paper.
\subsection{Overview of our approach} \label{sec:approach}
\begin{figure}[h!]
  \centering
  \includegraphics[width=0.38\textwidth]{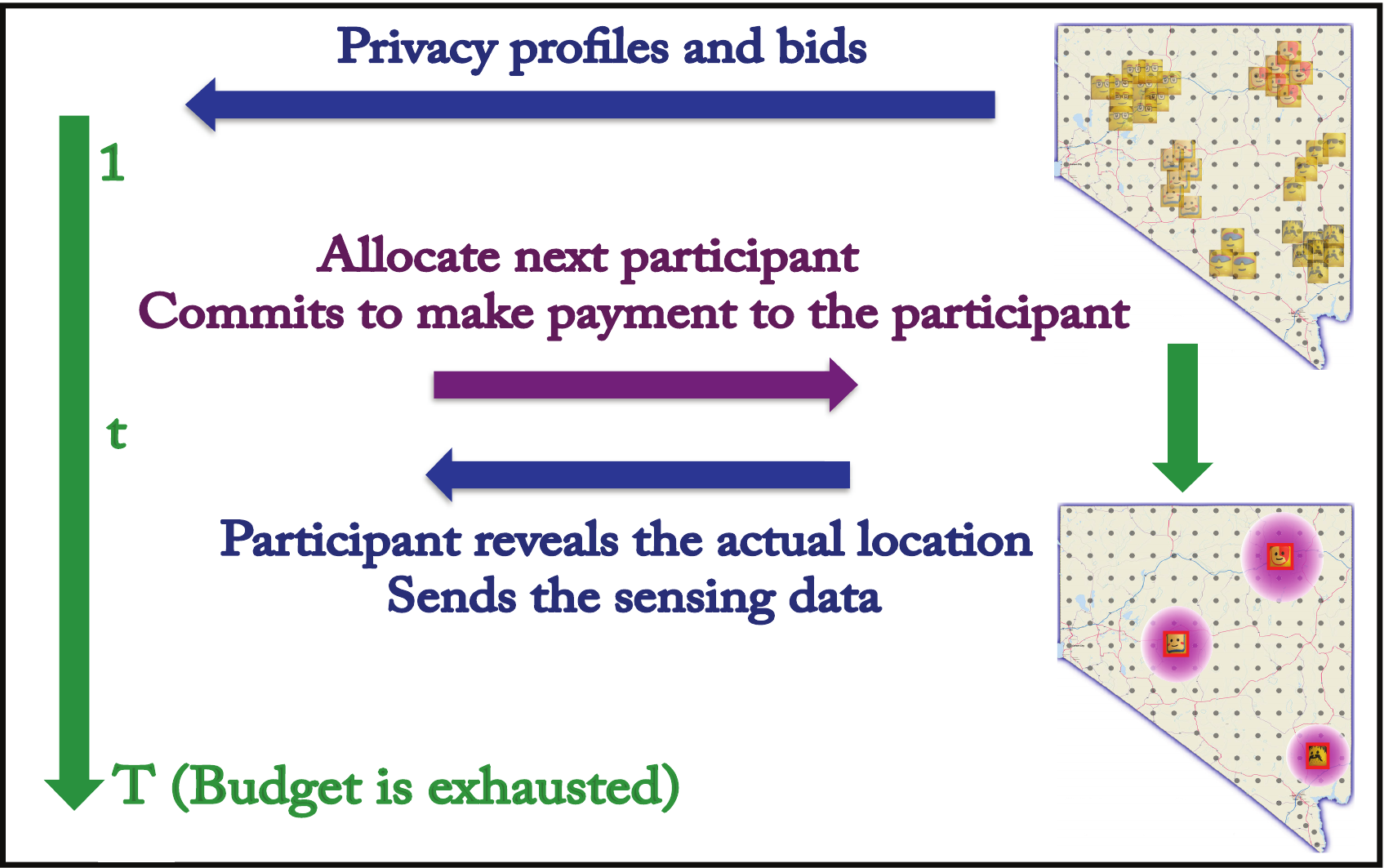}
  \caption{Illustration of the protocol by which the proposed system interacts with the users.}
  \label{fig:protocol}
\end{figure}

Our goal is to design policies for selecting (and compensating) the participants, which provide near-optimal utility for the sensing application under strict budget constraints. As basis for selection, the community sensing system receives obfuscated estimates of the private attributes. For concreteness, we focus on {\em sensor location} as private information, but our approach generalizes to other attributes.
The users also declare 
a bid or cost as the desired monetary incentive for participation and hence privacy tradeoff.
After receiving the bids, the mechanism sequentially selects a participant, commits to make her the payment, receives the actual private information, selects the next participant and so on. At the end, all selected participants are provided the agreed payment. Figure~\ref{fig:protocol} illustrates this protocol.

We model the participants as strategic agents who aim to maximize their profit, by possibly misreporting their private costs. As a consequence, we require the mechanism to be truthful. In order to capture a large class of sensing applications, we only require the utility function to satisfy submodularity, a natural diminishing returns condition 
\cite{1978-_nemhauser_submodular-max,
2007-aaai_krause_observation-selection}. 
To design our mechanism, we first reduce the sequential negotiation of the privacy tradeoff to the problem of adaptive submodular maximization \cite{2008-wine-journal-version_stochastic-submodular,2011-jair_krausea_adaptive-submodularity}. 
Then, we extend recent results on truthful budget feasible mechanisms for submodular functions \cite{2010-focs_singer_budget-feasible-mechanisms,2011-soda_improved-budget-feasible,2012-wsdm_singer_how-to-win-friends-and-influence-people} to the adaptive setting. 

Our main contributions are:
\begin{itemize}
	\item An integrated approach to community sensing by incentivizing users to share certain private information.
	\item A novel mechanism, \stgreedy, for budgeted recruitment of strategic participants, which achieves near-optimal utility for the community sensing application. The mechanism is general and of independent interest, suitable also for other applications, \emph{e.g.}, viral marketing.
%
	\item Evaluation of our approach on a realistic case study of air quality monitoring based on data obtained through Amazon Mechanical Turk \footnote{\url{https://www.mturk.com/mturk/}}.  
\end{itemize}
\subsection{Related Work} \label{sec:relatedwork}

\citet{2005-patent_schedule-based-ad-on-phone} propose to provide users with rewards such as free minutes to motivate them to accept mobile advertisements. 
\citet{2011-hci_targeted-ad-on-phone} develop MobiAd, a system for 
targeted mobile advertisements, 
by utilizing the rich set of information available on the phone
and suggesting the service providers to give discounts to the users, in order to incentivize use of the system.
\citet{2008-MSWiM_game-theory-location-sharing} propose a game theoretic model of privacy  for social networking-based mobile applications and presents a tit-for-tat mechanism by which users take decisions about their exposed location obfuscation for increasing personal or social utility. 
\citet{2012-pmc_trading-privacy-with-incentives} study a privacy game in mobile commerce, where users choose the degree of granularity at which to report their location and the service providers offer them monetary incentives under budget constraints. The best users' response and the optimal strategy for the company are derived by analyzing the Nash equilibrium of the underlying privacy game. This is very different from our setting as we focus on algorithmic aspects of the mechanism in choosing the best set of users for participation in community sensing. 
\citet{2012-comp-sust_incentive-schemes-communitysensing} and \citet{2012-internet-of-things_truthful-measurements} study the problem of incentivizing users in community sensing to report accurate measurements and place sensors in the most useful locations. While developing incentive-compatible mechanisms, they do not consider the privacy aspect.  \citet{2013-www_singla_truthful-incentives} develops online incentive-compatible and budget feasible mechanisms for procurement. However, they consider a simple modular utility function where each participant provides a unit value. This is not applicable to our community sensing setting which deals with more complex utility functions. \citet{2013-www_your-browsing-behavior-for-a-big-mac} study how users value their personally identifiable information  (PII) while browsing. The experiments demonstrate that users have different valuations, depending on the type and information content of private data. Higher valuations are chosen for offline PII, such as age and address, compared to browsing history. This work is complementary and supports the assertion that users indeed associate monetary valuations to certain private data.
\section{Problem Statement}\label{sec:model}
\begin{figure*}[t!]
\centering
   \subfigure[Population of users]{
     \includegraphics[width=0.25\textwidth]{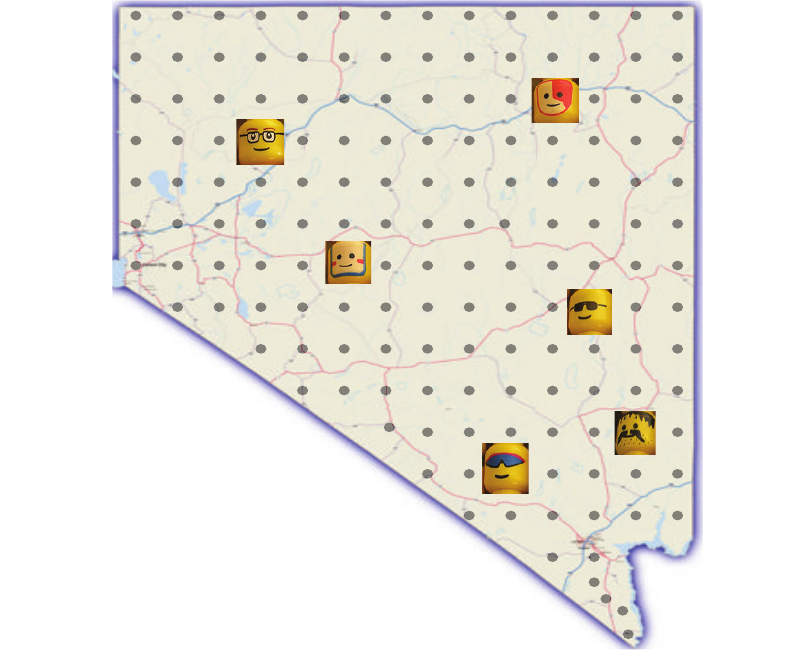}
     \label{fig:model-illustration-1}
   }
   \subfigure[Sensing profile of users]{
    \includegraphics[width=0.25\textwidth]{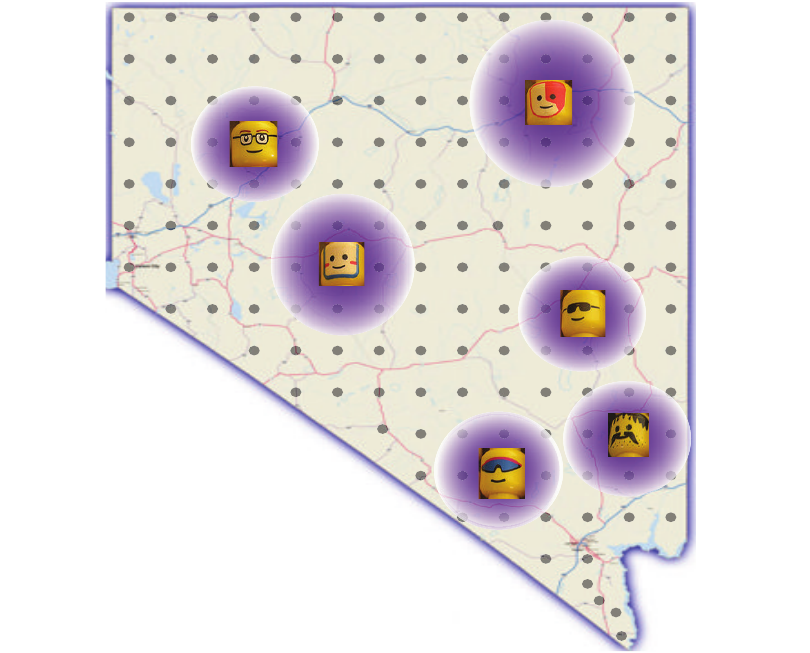}
    \label{fig:model-illustration-2}
   }
   \subfigure[Selected participants]{
     \includegraphics[width=0.25\textwidth]{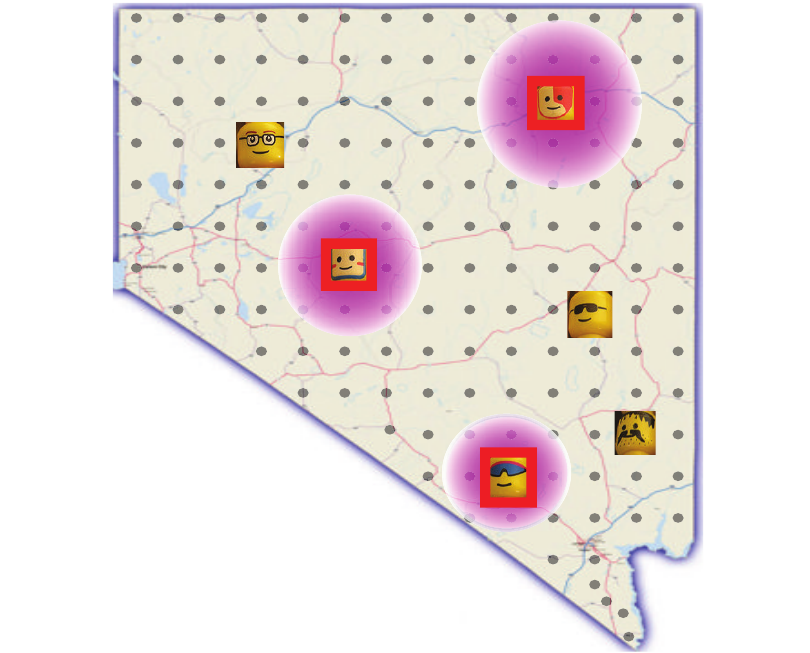}
     \label{fig:model-illustration-3}
   }
   \subfigure[Obfuscated user locations]{
     \includegraphics[width=0.25\textwidth]{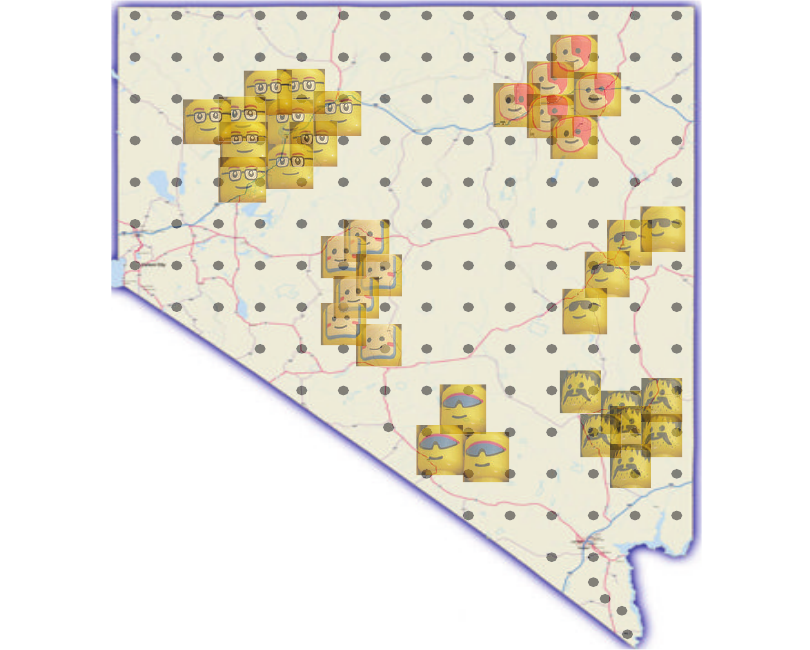}
     \label{fig:model-illustration-4}
   }
   \subfigure[Privacy profile of user \iw]{
    \includegraphics[width=0.25\textwidth]{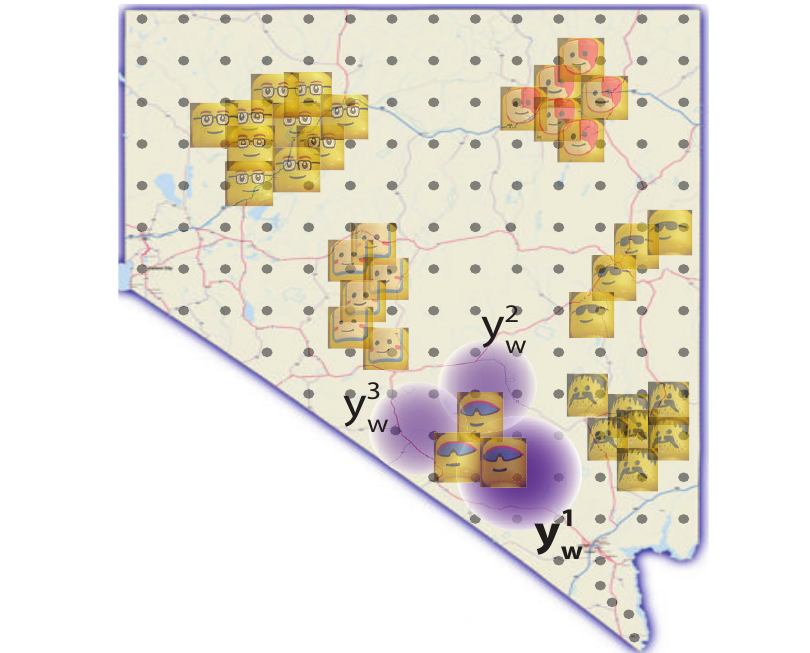}
    \label{fig:model-illustration-5}
   }
   \subfigure[Selection under uncertainty]{
     \includegraphics[width=0.25\textwidth]{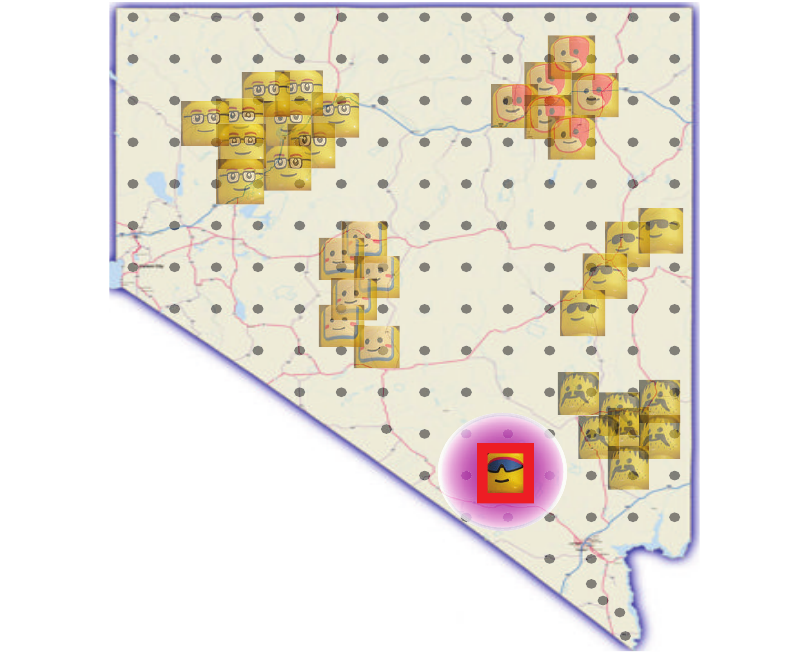}
     \label{fig:model-illustration-6}
   }
\caption{The sensing region is uniformly discretized into a set of locations $\icV$ indicated by the dots. ({\bf a}) illustrates a population of users, along with their sensing profiles in ({\bf b}). The set of users selected by the system in absence of privacy are shown in ({\bf c}). However, to protect privacy, users only share an obfuscated location with the system in ({\bf d}) and a collection of sensing profiles (\{$\iobs_{\iw}^1, \iobs_{\iw}^2$ and $\iobs_{\iw}^3$\} for user \iw)  in ({\bf e}). The privacy profile of user $\iw$, given by $\iObs_\iw$, is the uniform distribution over these sensing profiles, given by $P(\iObs_\iw=\iobs_{\iw}^i)=\frac{1}{3}$. ({\bf f}) shows the selection of the participants in presence of uncertainty introduced by privacy profiles. The actual sensing profile is only revealed to the system after a user has been selected.}
\label{fig:model-illustration}
\end{figure*}
We now formalize the problem addressed in this paper.
 
{\bf Sensing phenomena.} 
We focus on community sensing applications with the goal to monitor some spatial phenomenon, such as air quality or traffic. 
We discretize the environment as a finite set of locations $\icV$, where each  $\iv \in \icV$ could, \emph{e.g.}, denote a zip code 
or more fine grained street addresses, depending  on the application. We quantify the utility $\iobjf(\icA)$ of obtaining measurements from a set of locations $\icA$ using a set function $\iobjf:2^\icV\rightarrow\iReal$. Formally, we only require that $\iobjf$ is  \emph{nonnegative},  \emph{monotone} (i.e., whenever $\icA\subseteq\icA'\subseteq\icV$ it holds that $\iobjf(\icA)\leq \iobjf(\icA')$) and \emph{submodular}.
Submodularity is an intuitive notion of diminishing returns, stating that, for any sets $\icA\subseteq\icA'\subseteq\icV$, and any fixed location $\ia\notin\icA'$ it holds that $\iobjf(\icA\cup\{\ia\})-\iobjf(\icA)\geq \iobjf(\icA'\cup\{\ia\})-\iobjf(\icA')$.
As a simple, concrete example, we may derive some nonnegative value $\id_\ia$ for observing each location $\ia\in\icA$,
 and may define $\iobjf(\icA) = \sum_{\ia \in \icA} \id_\ia$. More generally, sensing at location $\ia\in\icV$ may actually cover a subset $\icS_\ia$ of nearby locations, and $\iobjf(\icA)=\sum\{\id_j: j\in\cup_{\ia\in\icA}\icS_\ia\}$.
These conditions are rather general, satisfied by many sensing utility functions 
and $f$ can capture much more complex notions, such as reduction of predictive uncertainty in a probabilistic model \cite{2007-aaai_krause_observation-selection}.
  

{\bf Sensing profile of users.} 
We consider a community $\icW$ of $|\icW|=N$ users, owning some sensing device such as a smartphone. Each user can make observations at a set of locations depending on her geolocation or mobility as well as the type of device used. 
We model this through a collection of \emph{sensing profiles} $\icO \subseteq 2^{\icV}$ whereby we associate each user $\iw \in \icW$ with a profile $\iobs_\iw \in \icO$, specifying the set of locations covered by her. This set $\iobs_\iw$ could be a singleton $\iobs_\iw=\{\ia\}$ for some $\ia\in\icV$, modeling the location of the user at a particular point in time, or could model an entire trajectory, visiting multiple locations in $\icV$.
We denote a given set of users $\icS \subseteq \icW$ jointly with their sensing profiles as ${\ibobs}_{\icS} \subseteq \icW \times \icO$. The goal is to select set of users $\icS$ (also called \emph{participants}) so as to maximize the 
utility of the sensing application given by $\iobjg({\ibobs}_{\icS}) = \iobjf(\icA) \text{ where } \icA = \bigcup_{\is \in \icS} \iobs_\is$. We assume that each user's maximal contribution to the utility is bounded by a constant $\ifmax$. 

{\bf Privacy profile of users.} 
\looseness -1 In order to protect privacy, we consider the setting where the exact sensing profiles $\iobs_\iw$ of the users (containing, \emph{e.g.}, tracks of locations visited) are not known to the sensing system. %
Instead, $\iobs_\iw$ is only shared after obfuscation with a random perturbation intended to reduce the risk of identifiability \cite{2002-journal-ufks_k-anonymity,2006-icalp_differential-privacy}.
The system's highly uncertain belief about the sensing profile of user $w$ can therefore be represented as 
a (set-valued) random variable (also called \emph{privacy profile}) $\iObs_\iw$ with $\iobs_\iw$ being its realization. For example, suppose $\iobs_\iw=\{\ia\}$ for some location $\ia$ (i.e., the user's private location is $\ia\in\icV$). In this case, the user may share with the system a collection of locations $\ia_1,\dots,\ia_m$ containing $\ia$ (but not revealing which one it is), w.l.o.g. $\ia=\ia_1$.  In this case the distribution shared $P(\iObs_\iw=\{a_i\})=\frac{1}{m}$ is simply the uniform distribution over the candidate locations. Figure~\ref{fig:model-illustration} illustrates the notions of sensing and privacy profiles for a  user.

We use $\ibObs_\icW = [\iObs_1,\dots,\iObs_N]$ to refer to the collection of all (independent) variables associated with population $\icW$ and assume that $\ibObs_\icW$ is distributed according to a factorial joint distribution $\iprobObs(\ibObs_\icW)=\prod_{\iw}P(\iObs_\iw)$. The sensing profile $\iobs_\iw$ (and the actual sensor data obtained from sensing at locations $\iobs_\iw$) is revealed to the application only after it commits to provide the desired incentives to the user $\iw$. 
Then, the goal is to select a set of users $\icS$ to maximize  
$\iExp_{\ibObs_\icW}[\iobjg(\ibobs_\icS)]$, \emph{i.e.}, the expected 
utility, where the expectation is taken over the realizations of 
$\ibObs_\icW$ \emph{w.r.t.} 
$\iprobObs(\ibObs_\icW)$. 

{\bf Incentive structure for privacy tradeoff.}
We assume that users are willing to share certain non-sensitive private information in return for monetary incentives. Each user $\iw$ has a private cost $\icost_\iw \in \iReal_{\ge0} $ that she experiences for her privacy tradeoff. Instead of revealing $\icost_\iw$, she only reveals a \emph{bid} $\ibid_\iw \in {\iReal_{\ge0}} $. We are interested in \emph{truthful} mechanisms, 
where it is a dominant strategy for a user to report $\ibid_\iw = \icost_\iw$, \emph{i.e.}, users cannot increase their profit (in \emph{expectation}) by lying about their true cost. We assume that costs have known bounded support, \emph{i.e.},  $\icost_\iw \in [\icmin,\icmax]$. 

{\bf Optimization problem.}
Given a strict budget constraint $\icB$, the goal of the sensing application is to design a mechanism $\iMechanism$, which implements an allocation policy to select participants $\icS$ and a payment scheme to make \emph{truthful} payments $\ipayment_\is$
to each of the participants, with the goal of maximizing the expected utility. Instead of committing to a fixed set of participants $\icS$ in advance (\emph{non-adaptive} policy), we are interested in mechanisms that implement an \emph{adaptive} policy taking into account the observations made so far (revealed sensing profiles of participants already selected) when choosing the next user. 
Formally, the goal of the mechanism 
is to adaptively select participants $\icS^*$ along with the payments $\ipayment_{\icS^*}$, such that
\begin{align}\label{eq:opt}
\icS^* &= \operatorname*{arg\,max}_{\icS \subseteq \icW} \iExp_{\ibObs_\icW}[\iobjg(\ibobs_\icS)]  \text{ subject to } \sum_{\is \in \icS} \ipayment_\is \leq \icB.
\end{align}
Here, the set of participants $\icS$ selected and the payments $\ipayment_\icS$ may depend on the realization of $\ibObs_\icW$ as well. We formally introduce adaptive policies in subsequent sections.

\section{Existing Mechanisms}\label{sec:algorithms}
We first review existing mechanisms that fall short of either privacy-preservation, adaptivity or truthfulness. In next section, we then build on these and present our main contribution: a privacy-respecting, truthful and adaptive mechanism.
\subsection{Non-private mechanisms} \label{sec:algorithms:noprivacy}
\looseness -1 Consider first an unrealistic setting, where the system has full information about the users' exact sensing profiles and their true costs. In such a setting, Problem~\ref{eq:opt} reduces to that of budgeted maximization of a monotone non-negative submodular function with non-uniform costs, studied by \citet{2004-operations_sviridenko_budgeted-submodular-max}. A simple algorithm combining partial enumeration with greedy selection 
guarantees a utility of at least $(1 - \sfrac{1}{e})$ $(=0.63)$ times that obtained by optimal selection {\opt}. This result is tight under reasonable complexity assumptions \cite{1998-_feige_threshold-of-ln-n}. 
We denote this setting and mechanism as $\greedy$. Note that each participant is paid their true cost in this untruthful setting. Now, consider the non-private setting with {\em unknown} true costs.
The problem then requires designing a truthful budget feasible mechanism for monotone submodular set functions, as done by \cite{2010-focs_singer_budget-feasible-mechanisms,2011-soda_improved-budget-feasible,2012-wsdm_singer_how-to-win-friends-and-influence-people}. In this setting, a constant factor $\sfrac{1}{7.91}$ $(=0.13)$ approximation compared to {\opt} can be achieved, using a mechanism that we will refer to as   
${\tgreedy}$. 
${\tgreedy}$ executes a greedy allocation on a reduced budget with carefully chosen stopping criteria (for ensuring budget feasibility), in order to select a set of participants and then computes the truthful payments to be made to them.
\begin{table}[t!]
\centering
\begin{tabular}{|l|c|c|}
\hline
& {\bf Untruthful} & {\bf Truthful} \\ \hline \hline
{\bf Priv. off} & {\small $\greedy$} & {\small $\tgreedy$} \\ \hline \hline
{\bf  Priv. on (Non-Ad.)} & {\small $\cgreedy$} & {\small $\ctgreedy$} \\ \hline
{\bf  Priv. on (Adaptive)} &  {\small $\sgreedy$} & {\small {\bf $\stgreedy$}}  \\ \hline
\end{tabular}
\caption{Different information settings and mechanisms.}\label{tab:algorithms:methodology}
\end{table}
\subsection{Non-adaptive mechanisms with privacy} \label{sec:algorithms:nonadaptive}
In our case, where privacy is preserved through random obfuscation, 
one must deal with the stochasticity caused by the uncertainty about users' sensing profiles.
Here, the objective 
\begin{align*}
G(\icS)\equiv\iExp_{\ibObs_\icW}[\iobjg(\ibobs_\icS)]=\sum_{\ibobs_\icW} P(\ibObs_\icW = \ibobs_\icW) f\left(\bigcup_{s\in\icS} \iy_s\right)
\end{align*}
in \eqref{eq:opt} can be seen as an expectation 
over multiple submodular set functions, one for each realisation of the privacy profile variables $\ibObs_\icW$.
However, 
as submodularity is preserved under expectations, the set function $G(\icS)$ is submodular as well.
One can therefore still apply the mechanisms $\greedy$ and ${\tgreedy}$ in order to obtain near-optimal \emph{non-adaptive} solutions (\emph{i.e.}, the set of participants is fixed in advance) to Problem~\eqref{eq:opt}.
We denote these non-adaptive (constant) mechanisms applied to our privacy-preserving setting as  ${\cgreedy}$ and ${\ctgreedy}$. 
\subsection{Untruthful, adaptive mechanisms with privacy} \label{sec:algorithms:adaptive}
\looseness -1 Instead of non-adaptively committing to the set $\icS$ of participants a priori, one may wish to obtain increased utility through adaptive (active/sequential) selection, \emph{i.e.}, by taking into account 
the observations from the users selected so far when choosing the next user. Without assumptions, computing such an optimal policy for Problem~\eqref{eq:opt} is intractable. 
Fortunately, as long as the sensing quality function $\iobjf$ is monotone and submodular,
 Problem~\eqref{eq:opt} satisfies a natural condition called \emph{adaptive submodularity} \cite{2011-jair_krausea_adaptive-submodularity}. This condition
generalizes the classical notion of submodularity to sequential decision / active selection problems as faced here.

%
{\bf Adaptive submodularity} requires, in our setting, that the expected benefit of any fixed user $\iw\in\icW$ given a set of observations (\emph{i.e.}, set of users and observed sensing profiles) can never increase as we make more observations. 
%
%
Formally, consider the \emph{conditional expected marginal gain} of adding a user $\iw \in \icW \setminus \icS$ to an existing set of observations $\ibobs_{\icS}\subseteq\icW\times\icO$:
\begin{align*}
\imarginal_{\iobjg}(w|\ibobs_\icS) = &\iExp_{\iObs_\iw}[\iobjg(\ibobs_{\icS} \cup \{(\iw,\iobs_\iw)\}) - \iobjg(\ibobs_{\icS}) | \ibobs_{\icS}] \quad \quad \\
\quad                  = \sum_{\iobs \in \icO} &\iP(\iObs_\iw = \iobs|\ibobs_{\icS}) \cdot [\iobjg(\ibobs_{\icS} \cup \{(\iw,\iobs)\}) - \iobjg(\ibobs_{\icS})].
\end{align*}
Function $g$ with distribution $\iprobObs(\ibObs_\icW)$ is {\em adaptive submodular}, if $\imarginal_{\iobjg}(w|\ibobs_\icS) \geq \imarginal_{\iobjg}(w|\ibobs_\icSalt) \text{ whenever } \ibobs_{\icS} \subseteq \ibobs_{\icSalt}$. Thus, the gain of a user $w$, in expectation over its unknown privacy profile, can never increase as we select and obtain data from more participants.
\begin{proposition}
Suppose $f$ is monotone and submodular. Then the objective $g$ and distribution $P$ used in Problem~\ref{eq:opt} are adaptive submodular.
\end{proposition}
Above Proposition follows from Theorem~6.1 
of \citet{2011-jair_krausea_adaptive-submodularity}, assuming distribution $P$ is factorial (i.e., the random obfuscation is independent between users). Given this problem structure,
for the simpler, untruthful setting (i.e., {\em known} true costs), we can thus use the sequential greedy policy for stochastic submodular maximization studied by \citet{2011-jair_krausea_adaptive-submodularity}. This approach is denoted by ${\sgreedy}$ and obtains a utility of at least
$(1 - \sfrac{1}{e})$  $(=0.63)$
times that of optimal sequential policy $\sopt$. 

Table~\ref{tab:algorithms:methodology} summarizes the settings and mechanisms considered so far.  They all fall short of at least one of the desired characteristics of privacy-preservation, truthfulness or adaptivity. In the next section, we present our main contribution -- {\stgreedy}, an adaptive mechanism for the realistic setting of privacy-sensitive and strategic agents. 




\section{Our main mechanism: \stgreedy}\label{sec:newalgorithms}
We now describe our mechanism $\iMechanism = (\iPolicy_\iMechanism, \iPayment_\iMechanism)$, with allocation policy $\iPolicy_\iMechanism$ and payment scheme $\iPayment_\iMechanism$. $\iMechanism$ first obtains the bids $\iBid_\icW$ and privacy profiles $P(\ibObs_\icW)$ from all users, runs the allocation policy $\iPolicy_\iMechanism$ to adaptively select participants $\icS$ and makes observations $\ibobs_\icS$ during selection. At the end, it computes payments $\ipayment_\icS$ using scheme $\iPayment_\iMechanism$. The allocation policy $\iPolicy_\iMechanism$ can be thought of as a decision tree. Formally, a policy $\ipolicy : 2^{\icW \times \icO} \rightarrow \icW$ is a partial mapping from observations $\ibobs_\icS$ made so far to the next user $\iw \in \icW \setminus \icS$ to be recruited, denoted by $\ipolicy (\ibobs_\icS) = \iw$. We seek policies that are provably competitive with the optimal (intractable) sequential policy $\sopt$. $\iPayment_\iMechanism$ computes payments which are truthful in expectation (a user cannot increase her total expected profit by lying about her true cost, for a fixed set of bids of other users) and individually rational ($\ipayment_\is \geq \ibid_\is$). For budget feasibility, the allocation policy needs to ensure that the budget $\icB$ is sufficient to make the payments $\ipayment_\icS$ to all selected participants. Next, we describe in detail the allocation policy 
and payment scheme 
of $\stgreedy$ with these desirable properties.

\subsection{Allocation policy of \stgreedy}  \label{sec:newalgorithms:allocation}
\looseness -1 Policy~\ref{alg:stgreedy} presents the allocation policy of $\stgreedy$. 
The main ingredient of the policy is to greedily pick the next user that maximizes the expected marginal gain $\imarginal_{\iobjg}(w|\ibobs_\icS)$ per unit cost.
The policy uses additional stopping criteria to enforce budget feasibility, similar to $\tgreedy$ \cite{2011-soda_improved-budget-feasible}. Firstly, it runs on a reduced budget $\sfrac{\icB}{\ialpha}$. Secondly, it uses a proportional share rule ensuring that the expected marginal gain per unit cost for the next potential participant is at least equal to or greater than the expected utility of the new set of participants divided by the budget.
We shall prove below that $\ialpha = 2$ achieves the desired properties.


\subsection{Payment characterization of \stgreedy}  \label{sec:newalgorithms:payments}
The payment scheme is based on the characterization of threshold payments used by $\tgreedy$ \cite{2010-focs_singer_budget-feasible-mechanisms}. However, a major difficulty arises from the fact that the computation of payments for a participant depends also on the unallocated users, whose sensing profiles are not known to the mechanism. Let $\icS$ denote the set of participants allocated by $\iPolicy_\iMechanism$ along with making observations $\ibobs_\icS$. Let us consider the set of all possible realizations of $\ibObs_\icW = \ibobs_\icW \subseteq \icW \times \icO$ consistent with $\ibobs_\icS$, i.e., $\ibobs_\icS \subseteq \ibobs_\icW$. We denote this set by $\ibobsSet_{\icW, \icS} = [\ibobs^1, \ibobs^2 \dots \ibobs^\ir \dots \ibobs^\iZ]$, where $\iZ = |\ibobsSet_{\icW, \icS}|$. We first discuss how to compute the payment for each one of these possible realizations $\ibobs^\ir \in \ibobsSet_{\icW, \icS}$, denoted by $\ipaymentconst_{\is}(\ibobs^\ir)$ (where $d$ indicates here an association with the deterministic setting of knowing the exact sensing profiles of all users $\iw \in \icW$). These payments for specific realizations are then combined together to compute the final payment to each participant.

\begin{algorithm}[t!]
\nl {\bf Input}: \emph{budget}~$\icB$; \emph{users}~$\icW$; \emph{privacy profiles}~$\ibObs_\icW$; \emph{bids}~$\iBid_\icW$; \emph{reduced budget factor}~$\ialpha$\;
\nl {\bf Initialize}:\\
    \begin{itemize}
    \item {\bf Outputs}: \emph{participants}~$\icS \leftarrow \emptyset$; \emph{observations}~$\ibobs_\icS \leftarrow \emptyset$; \emph{marginals}~$\iMarginal_\icS \leftarrow \emptyset$;
    \item {\bf Variables}: \emph{remaining users}~$\icWalt \leftarrow \icW$;
    \end{itemize}
\Begin{
\nl  \While{$\icWalt \neq \emptyset$}{
\nl		  $\iwstar \leftarrow \operatorname*{arg\,max}_{\iw \in \icWalt} \frac{\imarginal_{\iobjg}(\iw|\ibobs_\icS)}{\ibid_\iw}$ \; 
\nl               $\imarginal_{\iwstar} \leftarrow \imarginal_{\iobjg}(\iwstar|\ibobs_\icS)$ \; 
\nl              \If{$\iBid_\icS + \ibid_\iwstar \leq \icB$}{
\nl                  \If{$\ibid_\iwstar \leq \frac{\icB}{\ialpha} \cdot \frac{\imarginal_{\iwstar}}{\big((\sum_{\is \in \icS} \imarginal_{\is}) + \imarginal_{\iwstar}\big)}$}{
\nl                     $\icS \leftarrow \icS \cup \{\iwstar\}$; $\iMarginal_{\icS} \leftarrow \iMarginal_{\icS} \cup \{{\imarginal_{\iwstar}}\}$ \; 
\nl                     Observe $\iobs_{\iwstar}$ ;  $\ibobs_{\icS} \leftarrow \ibobs_{\icS} \cup \{(\iwstar, \iobs_{\iwstar})\}$;
\nl             		    $\icWalt \leftarrow \icWalt \setminus \{\iwstar\}$ \;
                     }
\nl                  \Else{
\nl                      $\icWalt \leftarrow \emptyset$ ;
                    }
                }
\nl             \Else{
\nl             		$\icWalt \leftarrow \icWalt \setminus \{\iwstar\}$ \;
              	}
    }
}
\nl {\bf Output}: $\icS$; $\ibobs_\icS$; $\iMarginal_\icS$\\
\caption{Allocation policy of \stgreedy}
\label{alg:stgreedy} 
\end{algorithm}
{\bf Payment $\ipaymentconst_{\is}$ for a given $\ibobs_\icW$.} Consider 
the case where the variables $\ibObs_\icW$ are in state $\ibobs_\icW\in\ibobsSet_{\icW, \icS}$ and let $\icS$ be the set of participants allocated by the policy. 
We use the well-known characterization of \citet{1981-mor_myerson_optimal-auction-design} of truthful payments in single-parameter domains. It states that a mechanism is truthful if \emph{i)} the allocation rule is monotone (\emph{i.e.}, an already allocated user cannot be unallocated by lowering her bid, for a fixed set of bids of others) and \emph{ii)} allocated users are paid threshold payments (\emph{i.e.}, the highest bid they can declare before being removed from the allocated set). Monotonicity follows naturally from the  greedy allocation policy, which sorts users based on expected marginal gain per unit cost. To compute threshold payments, we need to consider a maximum of all the possible bids that a user can declare and still get allocated. We next explain how this can be done. 

Let us renumber the users $\icS = \{1,\dots,\ii,\dots,\ik\}$ in the order of their allocation. 
and let us analyze the payment for participant $\is = \ii$. Consider running the policy on an alternate set $\icWalt = \icW \setminus \{\ii\}$ and let $\icSalt = \{1,\dots,\iij,\dots,\ikalt\}$ be the allocated set (users renumbered again based on order of allocation when running the policy on $\icWalt$). $\iMarginal_\icS$ and $\iMarginalalt_\icSalt$ are the marginal contributions of the participants in the above two runs of the policy. We define $\imarginal_{\ii(\iij)}$ to be the marginal contribution of $\ii$ (from $\icS$) if it has to replace the position of $\iij$ (in set $\icSalt$). Now, consider the bid that $\ii$ can declare to replace $j$ in $\icSalt$ by making a marginal contribution per cost higher than $\iij$, given by $\ibid_{\ii(\iij)} = \frac{\imarginal_{\ii(\iij)} \cdot \ibid_\iij}{\imarginalalt_\iij}$. Additionally, the bid that $\ii$ can declare must satisfy the proportional share rule, denoted by $\irho_{\ii(\iij)} = \frac{\icB}{\ialpha} \cdot \sfrac{\imarginal_{\ii(\iij)}}{\big((\sum_{\isalt \in [\iij-1]} \imarginalalt_{\isalt}) + \imarginal_{\ii(\iij)}\big)}$. By taking the minimum of these two values, we get $\ipaymentconst_{\ii(\iij)} = \min(\ibid_{\ii(\iij)}, \irho_{\ii(\iij)})$ as the bid that $\ii$ can declare to replace $\iij$ in $\icSalt$. The threshold payment for participant $\is = \ii$ is given by $\ipaymentconst_{\ii} = \max_{j \in [\ikalt + 1]}\ipaymentconst_{\ii(\iij)}$.


{\bf Computing the final payment $\ipayment_{\is}$.} For each $\ibobs^\ir \in \ibobsSet_{\icW, \icS}$, compute $\ipaymentconstr_{\ii} = \ipaymentconst_{\ii}(\ibobs^\ir)$.
The final payment made to participant $\is$ is given by $\ipayment_\is = \sum_{\ibobs^\ir \in \ibobsSet_{\icW, \icS}} \iP(\ibObs_\icW = \ibobs^\ir|\ibobs_{\icS}) \cdot \ipaymentconstr_{\is}$.
Note that the set $\ibobsSet_{\icW, \icS}$ could be exponentially large, and hence computing the exact $\ipayment_{\is}$ may be intractable. However, one can use sampling to get estimates of $\ipayment_{\is}$ in polynomial time (using Hoeffding's inequality to bound sample complexity) and thus implement an approximately truthful payment scheme to any desired accuracy. Further, note that the approximation guarantees of $\iMechanism$ do not require computation of the payments at all, and only require execution of the allocation policy, which runs in polynomial time.

\subsection{Analysis of \stgreedy}  \label{sec:newalgorithms:theory}
We now analyze the mechanism and prove its desirable properties.
The proofs of all theorems are presented in the extended version of the paper \cite{2013-arxiv_singla_incentives-privacy}. We only sketch them here.
\begin{theorem}\label{theorem:truthful}
${\stgreedy}$ is truthful in expectation, i.e., no user can increase her profit in expectation by lying about her true cost, for a fixed set of bids of other users.
\end{theorem}
Firstly, truthfulness of payments $\ipaymentconstr_{\is}$ is proved for a considered realization $\ibobs^\ir$. This is done by showing the monotonicity property of the greedy allocation policy and proving the threshold nature of the payment $\ipaymentconstr_{\is}$. Truthfulness of the actual payment $\ipayment_{\is}$ follows from the fact that it is a linear combination of individually truthful payments $\ipaymentconstr_{\is}$.
\begin{theorem}\label{theorem:rational}
Payments made by ${\stgreedy}$ are individually rational, i.e. $\ipayment_\is \geq \ibid_\is$.
\end{theorem}
This is proved by showing a lower bound of $\ibid_\is$ on each of the payments $\ipaymentconstr_{\is}$ used to compute the final payment $\ipayment_{\is}$.
\begin{theorem}\label{theorem:budgetfeasible}
For $\alpha = 2$, ${\stgreedy}$ is budget feasible, i.e., $\ipayment_\icS \leq \icB$. Moreover, an application specific tighter bound on $\alpha$ can be computed to better utilize the budget.
\end{theorem}
We first show that when full budget $\icB$ is used by mechanism, the maximum raise in bid $\ibidalt_\is$ that a participant $\is$ can make, keeping the bids of other users to be the same,  to still get selected by mechanism is upper-bounded by $\ialpha \cdot \icB \cdot \sfrac{\imarginal_\is}{(\sum_{\is' \in \icS} \imarginal_{\is'})}$. By adapting the proof of \citet{2011-soda_improved-budget-feasible}, we prove that $\alpha$ is bounded by $2$. Surprisingly, this payment bound on $\alpha$ holds irrespectively of the payment scheme used by the mechanism. 
Hence, when the budget is reduced by $\alpha = 2$, this results in an upper bound on the payments made to any participant by $\icB\cdot\sfrac{\imarginal_\is}{(\sum_{\is' \in \icS} \imarginal_{\is'})}$. Summing over these payments ensures budget feasibility.
Moreover, by adapting a proof from \citet{2010-focs_singer_budget-feasible-mechanisms}, we show that a tighter bound on $\alpha$ can be computed based on the characterization of threshold payments used by $\stgreedy$. Intuitively, the proof is based on the fact that a raise in bid that a participant can make depends on how much utility the application would lose if she refused to participate.
\begin{theorem}\label{theorem:approximation}
For $\alpha = 2$, ${\stgreedy}$ achieves a utility of at least $\Bigl(\frac{e - 1}{3e}-\gamma\Bigr)$ times that obtained by the optimal  policy $\sopt$ with full knowledge of the true costs. Hereby, $\gamma$ is the ratio of the participants' largest marginal contribution $\ifmax$ and the expected utility achieved by \sopt.
\end{theorem}
\noindent We show that, because of the diminishing returns property of the utility function, the stopping criteria used by the mechanism based on proportional share and using only an $\ialpha$ proportion of the budget still allows the allocation of sufficiently many participants to achieve a competitive amount of utility. 
As a concrete example, if each participant can contribute at most 1\% to the optimal utility (i.e., $\gamma=0.01$), Theorem~\ref{theorem:approximation} guarantees a constant  approximation factor of $0.20$.


\section{Experimental Evaluation}\label{sec:experiments}
In this section, we carry out extensive experiments to understand the practical performance of our mechanism on a realistic community sensing case study. 

\begin{figure*}[]
\centering
   \subfigure[Bids (\$) and Sensitivity]{
     \includegraphics[width=0.32\textwidth]{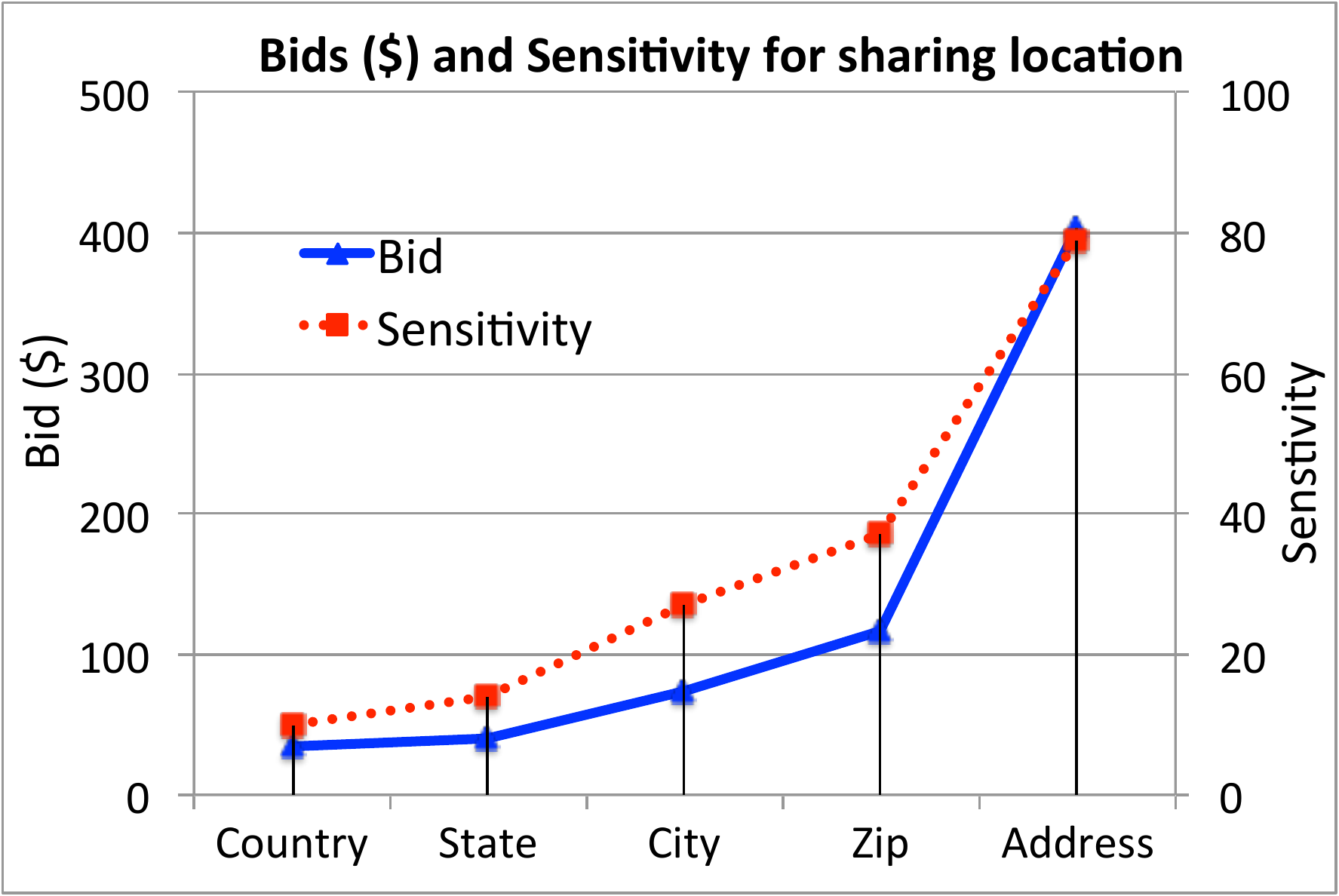}
     \label{fig:bids-senstivity}
   }
   \subfigure[Distributions of Zip-Bids]{
     \includegraphics[width=0.32\textwidth]{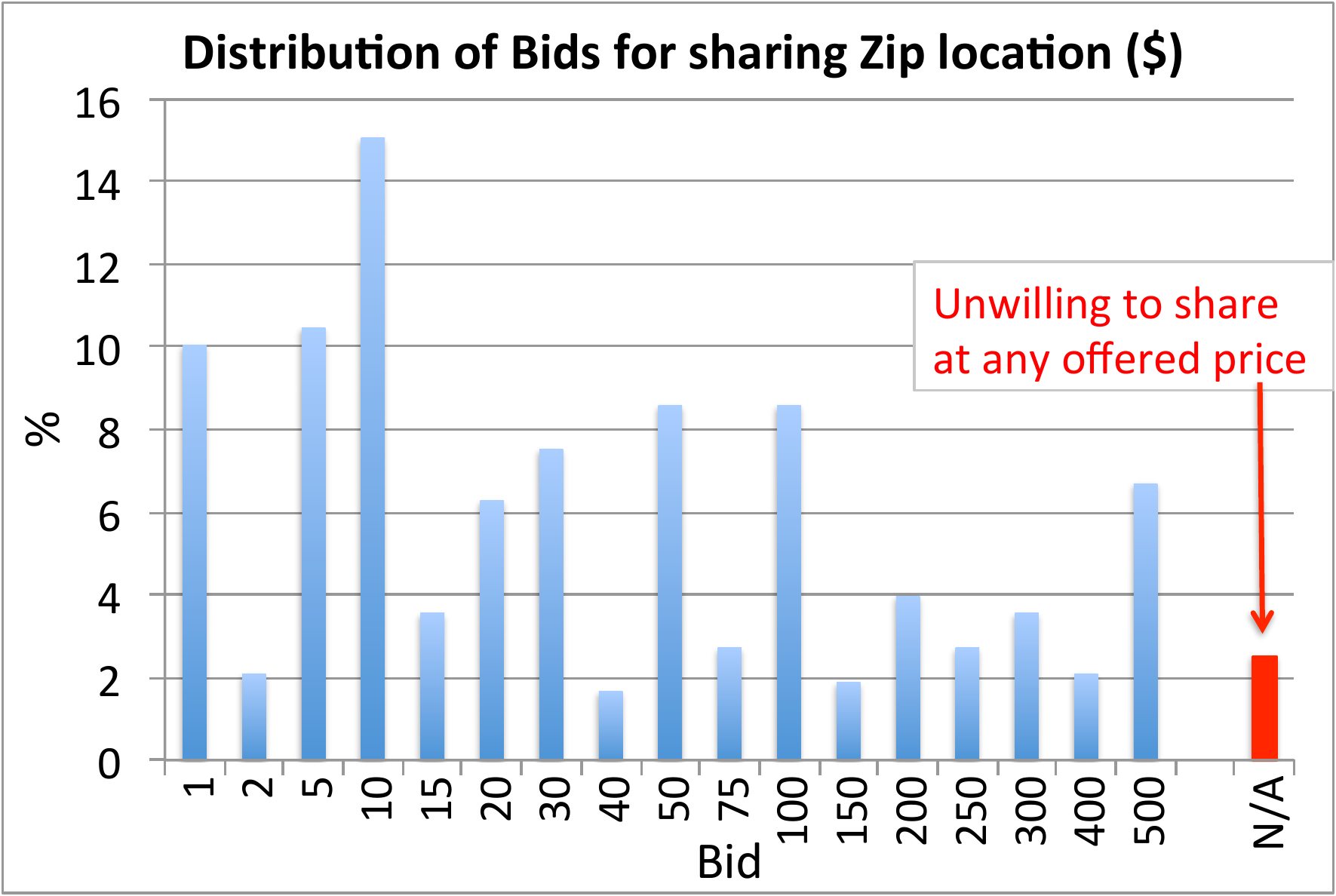}
     \label{fig:bids-distribution}
   }
   \subfigure[Corr. of Zip-Bids vs Mobility (miles)]{
     \includegraphics[width=0.32\textwidth]{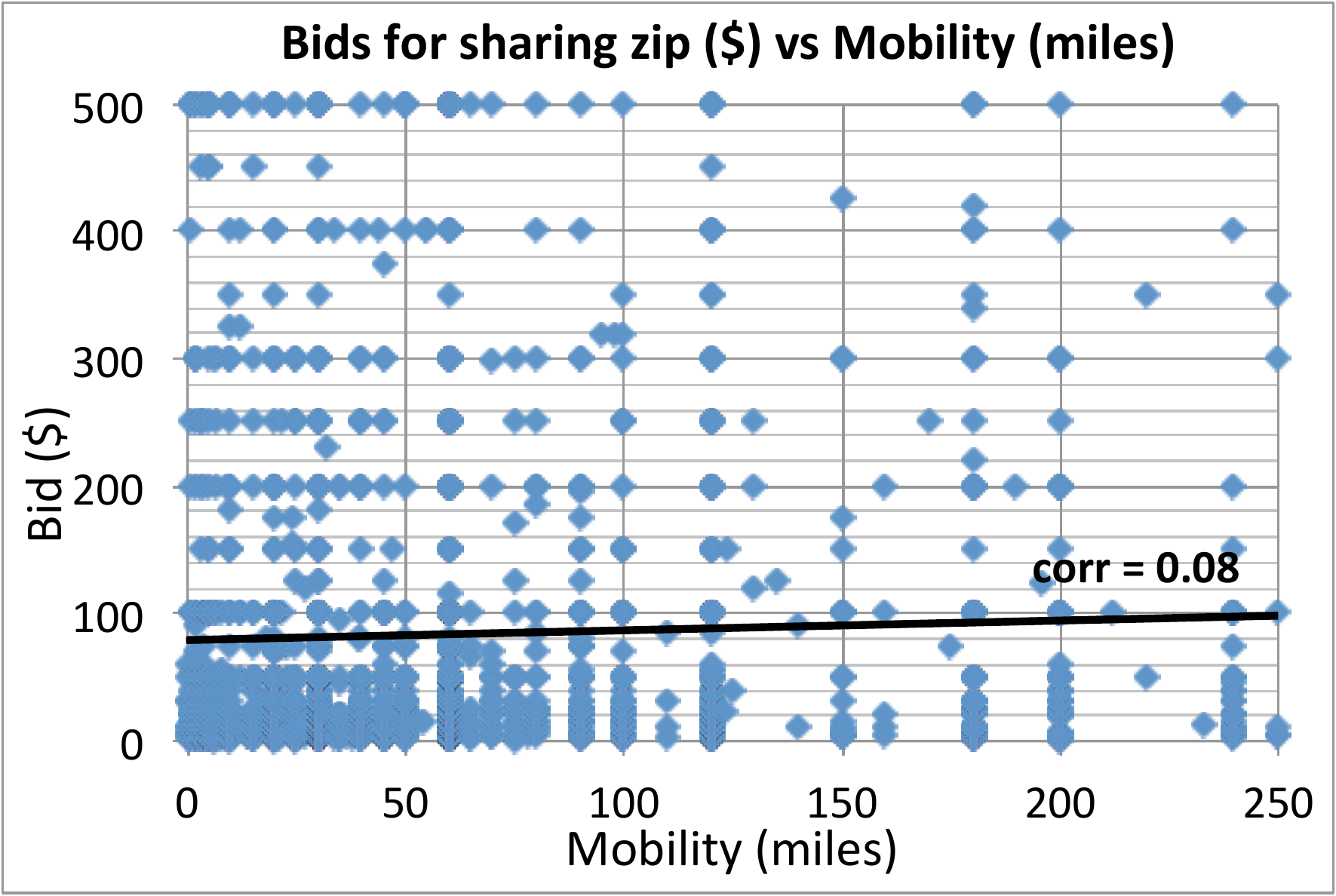}
     \label{fig:correlation-bidzip-daydistance}
   }
\caption{ ({\bf a}) Bids (\$) and sensitivity ([1-100]) for different levels of privacy tradeoff; ({\bf b}) Distribution of bids (\$) for sharing location at a granularity level of zip codes;  ({\bf c}) Correlation of bids (\$) (for sharing zip) with mobility (daily distance in miles).}
\label{fig:amt-stats}
\end{figure*}


{\bf Benchmarks.} We compare 
against the following benchmarks and state-of-the-art mechanisms.
\begin{itemize}
\item \sgreedy (unrealistically) assumes access to the true costs of the users, thus measuring the loss incurred by  \stgreedy for enforcing truthfulness and serving as upper bound benchmark on untruthful mechanisms.
\item \random  allocates users randomly until the budget is exhausted and pays each participant its true cost. 
This represents a lower bound benchmark on untruthful mechanisms.
\item \ctgreedy is the non-adaptive variant of \stgreedy and the state-of-the-art  truthful mechanism.
\item \tgreedy (unrealistically) assumes access to the exact sensing profiles of the users and hence provides insights in measuring the loss incurred 
due to privacy protection. 
\end{itemize}

{\bf Metrics and experiments.} 
The primary metric we measure is the utility acquired by the application. 
We also measure 
budget required 
to achieve a specified utility. To this end, we conduct experiments by varying the given budget and then varying the specified utility, for a fixed obfuscation level. To further understand the impact of random obfuscation, we then vary the level of obfuscation
and measure i) \emph{\%~Gain from adaptivity} (\stgreedy vs. \ctgreedy), ii) \emph{\% Loss from truthfulness} (\stgreedy vs. \sgreedy), and iii) \emph{\% Loss from privacy} (\stgreedy vs. \tgreedy).
We present below the results obtained based on data gathered from Mechanical Turk (henceforth MTurk). The primary purpose of using Mechanical Turk (MTurk) data is to evaluate on realistic distributions rather than making assumptions about bids and participants' mobility. We carried out experiments on simulated distributions as well with qualitatively similar results. 
\subsection{Experimental setup and data sets}\label{sec:experiments:datasets}
We now describe our setup and data collection from MTurk.

{\bf Community sensing application.} 
Suppose we wish to monitor air quality using mobile sensors \cite{2010-iwgs_opensense}. We consider a granularity level of zip codes and locations $\icV$ correspond to the zip codes of state Nevada, USA. We obtained information related to latitude, longitude, 
city and county  of these zips from publicly available data  \footnote{\url{http://www.populardata.com/downloads.html}}. This represents a total of 220 zip codes located in 98 cities and 17 counties. 
In order to encourage spatial coverage, we choose our objective $\iobjf$ such that one unit utility is obtained for every zip code location observed by the selected participants.
To simulate a realistic population of the $N$ users,
we also obtained the population statistics for these zip codes \footnote{\url{http://mcdc2.missouri.edu/}}.

{\bf MTurk data and user attributes.}
We posted a Human Intelligence Task (HIT) on MTurk in form of a survey, where workers were told about an option to participate in a community sensing application. Our HIT on MTurk clearly stated the purpose as purely academic, requesting workers to provide correct and honest information. 
The HIT presented the application scenario and asked workers about their willingness ("yes/no") to participate in such applications. 75\% (487 out of 650) responded positively. Workers were asked to express their sensitivity (on scale of [1-100]), as well as the payment bids (in range of [1-500] \$) they desire to receive about exposing their location at the granularity of home address, zip, city, state or country respectively. 
Additionally, workers were asked about their daily mobility to gather data for defining the sensing radii of the users in our experiments. 

A total of 650 workers participated in our HIT, 
restricted to workers from the USA with more than 90\% approval rate and were paid a fixed amount each. 
We used the data of 487 workers for our experiments, who responded positively to participate in the envisioned application.
Figure~\ref{fig:bids-senstivity} shows the mean bids and expressed sensitivity for different levels of obfuscation. Figure~\ref{fig:bids-distribution} shows the distribution of bids for exposing zip level location information. A mean daily mobility of 18 miles was reported. Figure~\ref{fig:correlation-bidzip-daydistance} shows 
no correlation between their daily mobility (related to user's sensing radius and hence utility) and bids for exposing zip code information (related to user's bid).

{\bf Parameter choices and user profiles.} 
We consider a population of size $N=500$, distributed according to the population statistics for the zip codes. We used the distribution of bids reported for sharing location at a granularity level of zip codes. We set $\icmin$ = 0.01 and $\icmax$ = 1 by scaling the bids in this range. For a given location of a user, we used the distributions of daily mobility to define the sensing radius of the users. We set the maximum possible utility obtained from each user to $\ifmax=15$ by limiting the maximal number of observable zip code locations of each user to 15, which are randomly sampled from the locations covered by the user's sensing radius. 

\looseness -1 Given a user's zip location, the sensing profile of the user is uniquely specified. To create privacy profiles, we used obfuscated user locations, by considering obfuscation at city or state level in which the user is located. We also considered obfuscation within a fixed radius, centered around the user's location. For each of the obfuscated zip codes, multiple corresponding sensing profiles are generated, which collectively define the user's privacy profile.

\subsection{Results}\label{sec:experiments:results}
We now discuss the findings from our experiments.

{\bf Computing tighter bounds on payment.}
Based on Theorem~\ref{theorem:budgetfeasible}, we compute tighter bounds on the payment and optimized the budget reduction factor $\alpha$ used by our mechanism in an application specific manner. 
\begin{figure}[h!]
\centering
   \subfigure[Acquired utility]{
     \includegraphics[width=0.22\textwidth]{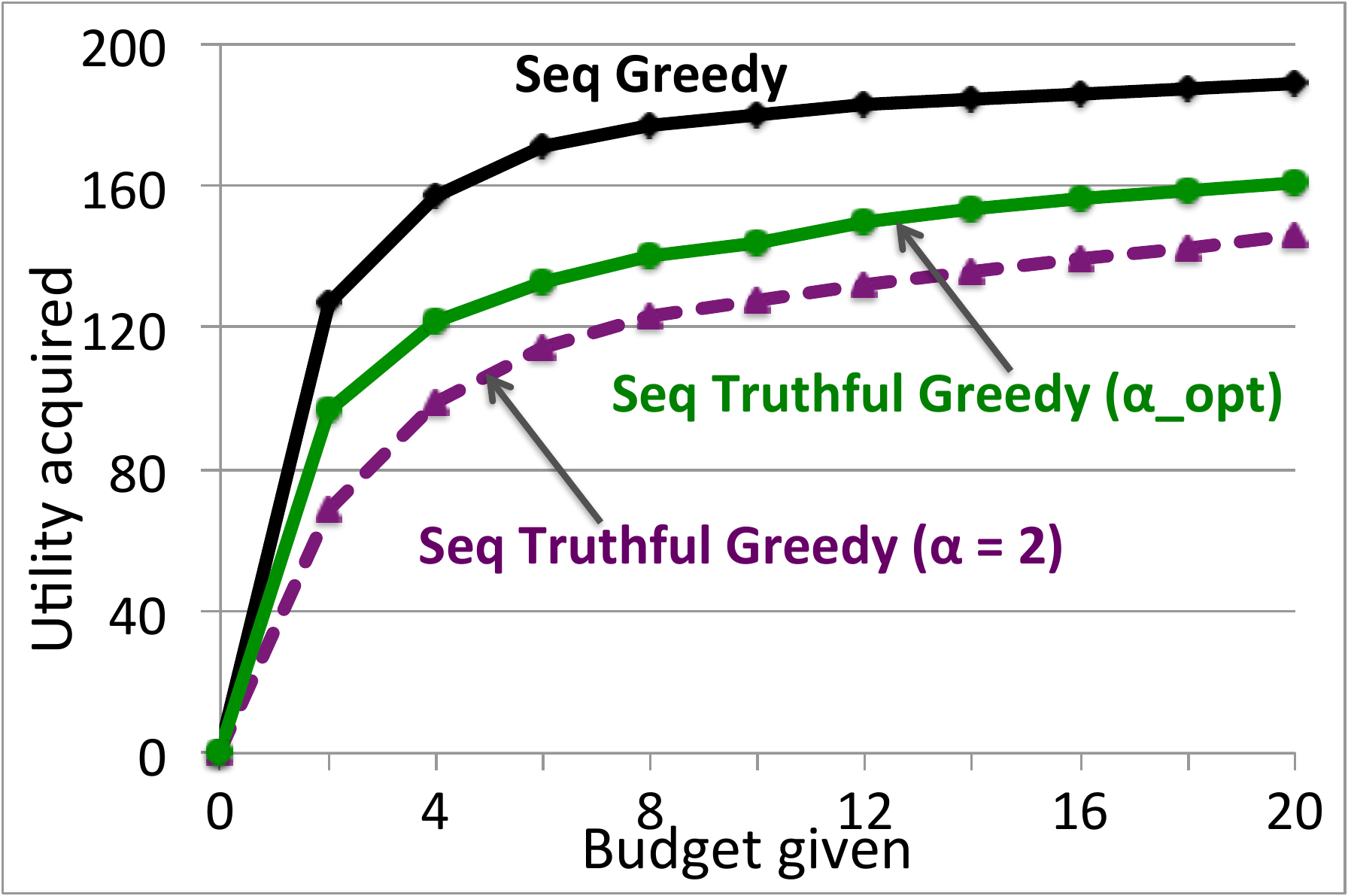}
     \label{fig:bids-senstivity}
   }
   \subfigure[Budget required]{
     \includegraphics[width=0.22\textwidth]{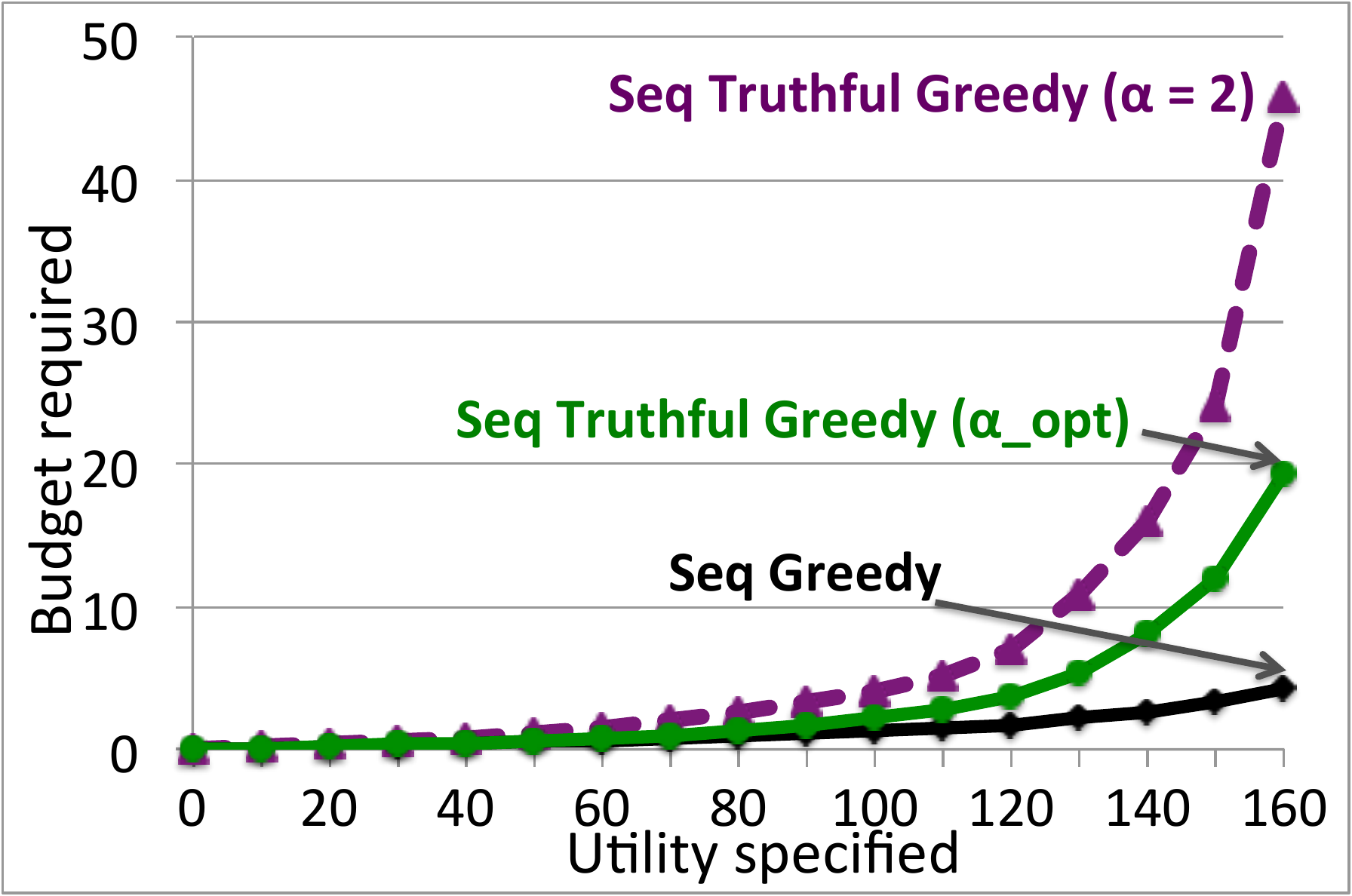}
     \label{fig:budget_vary-alpha}
   }
\caption{({\bf a}) and ({\bf d}) compares  \stgreedy using $\alpha = 2$ \emph{w.r.t.} to a variant using an optimized value of $\alpha$.}
\label{fig:budget_vary-alpha}
\end{figure}
\begin{figure*}[t!]
\centering
   \subfigure[Utility: Varying given budget]{
     \includegraphics[width=0.32\textwidth]{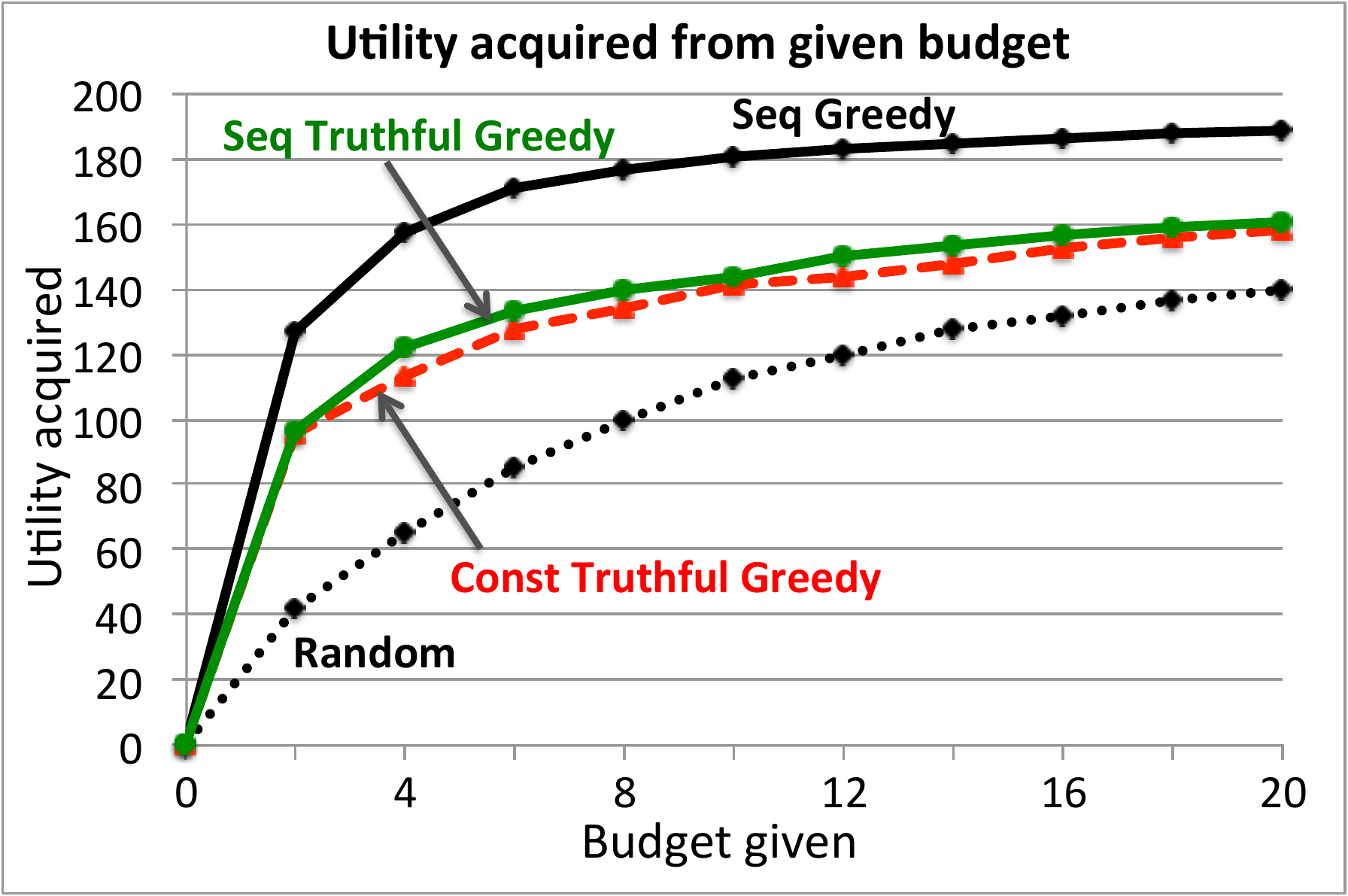}
     \label{fig:utility_vary-budget}
   }
   \subfigure[Utility: Varying obfuscation]{
    \includegraphics[width=0.32\textwidth]{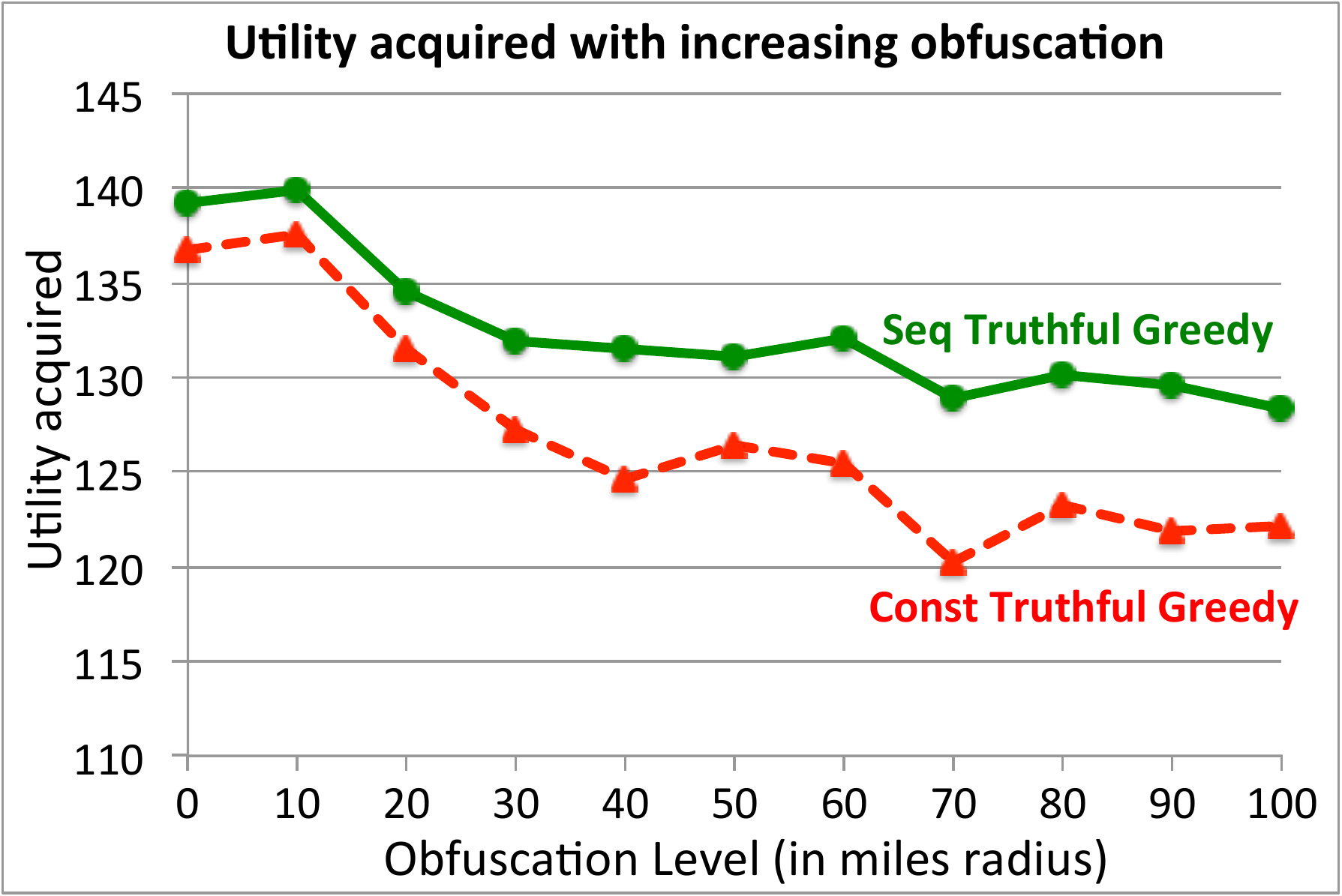}
    \label{fig:utility_vary-obfuscation}
   }
   \subfigure[\% Utility change: Varying obfuscation]{
     \includegraphics[width=0.32\textwidth]{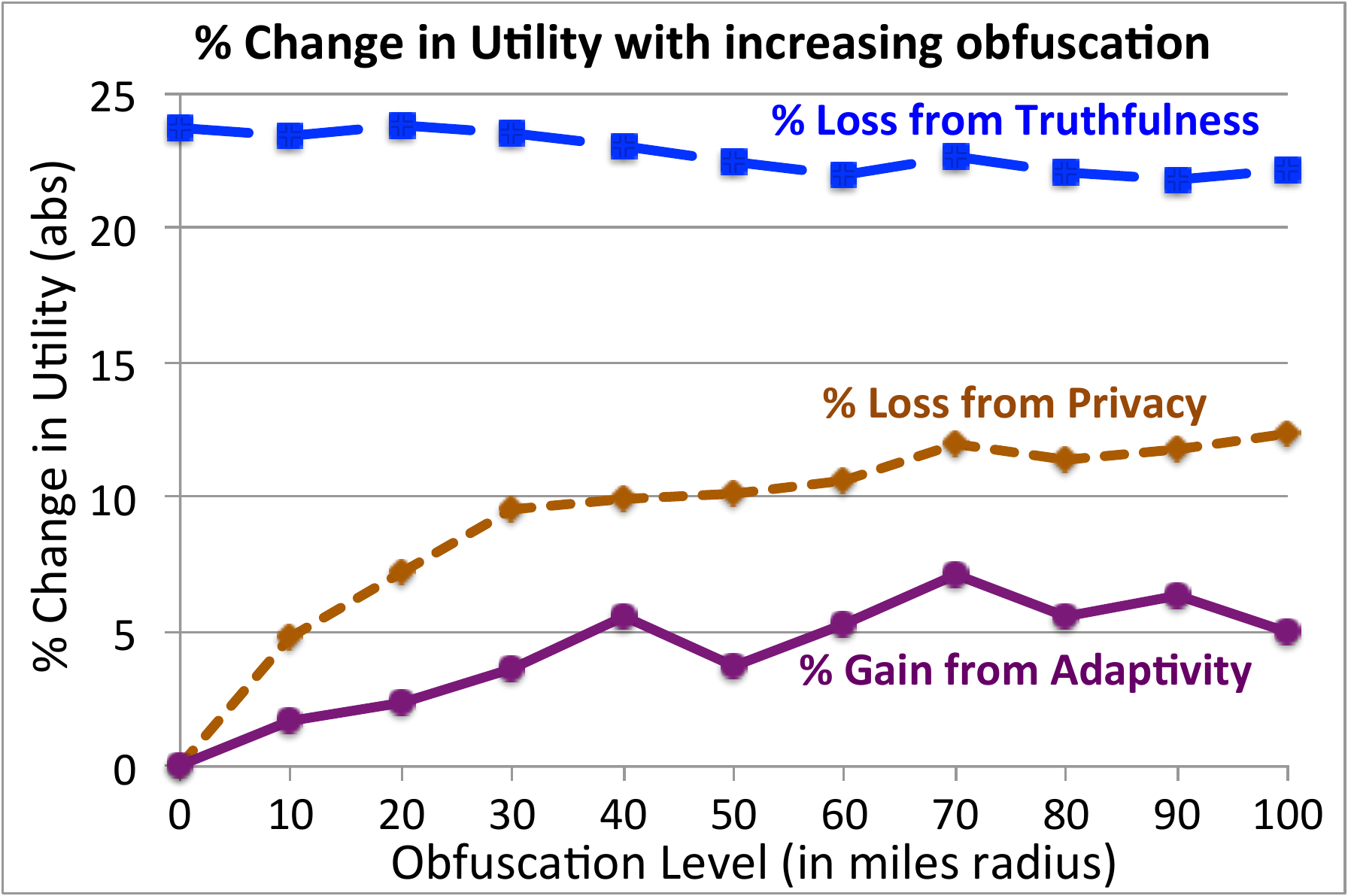}
     \label{fig:utility-change_vary-obfuscation}
   }\\[-2mm]
   \subfigure[Budget (\$): Varying specified utility]{
     \includegraphics[width=0.32\textwidth]{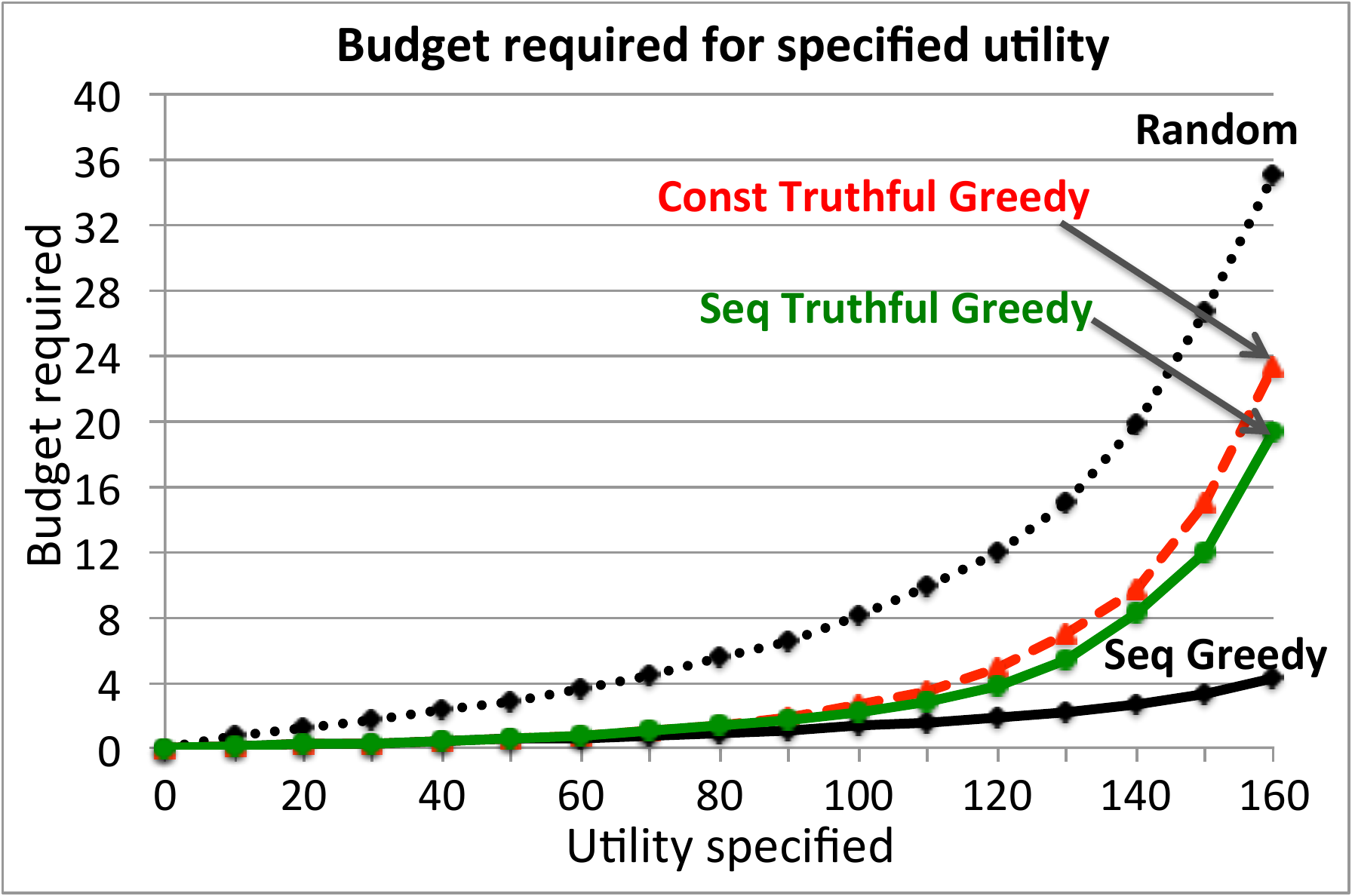}
     \label{fig:budget_vary-utility}
   }
   \subfigure[Budget (\$): Varying obfuscation]{
    \includegraphics[width=0.32\textwidth]{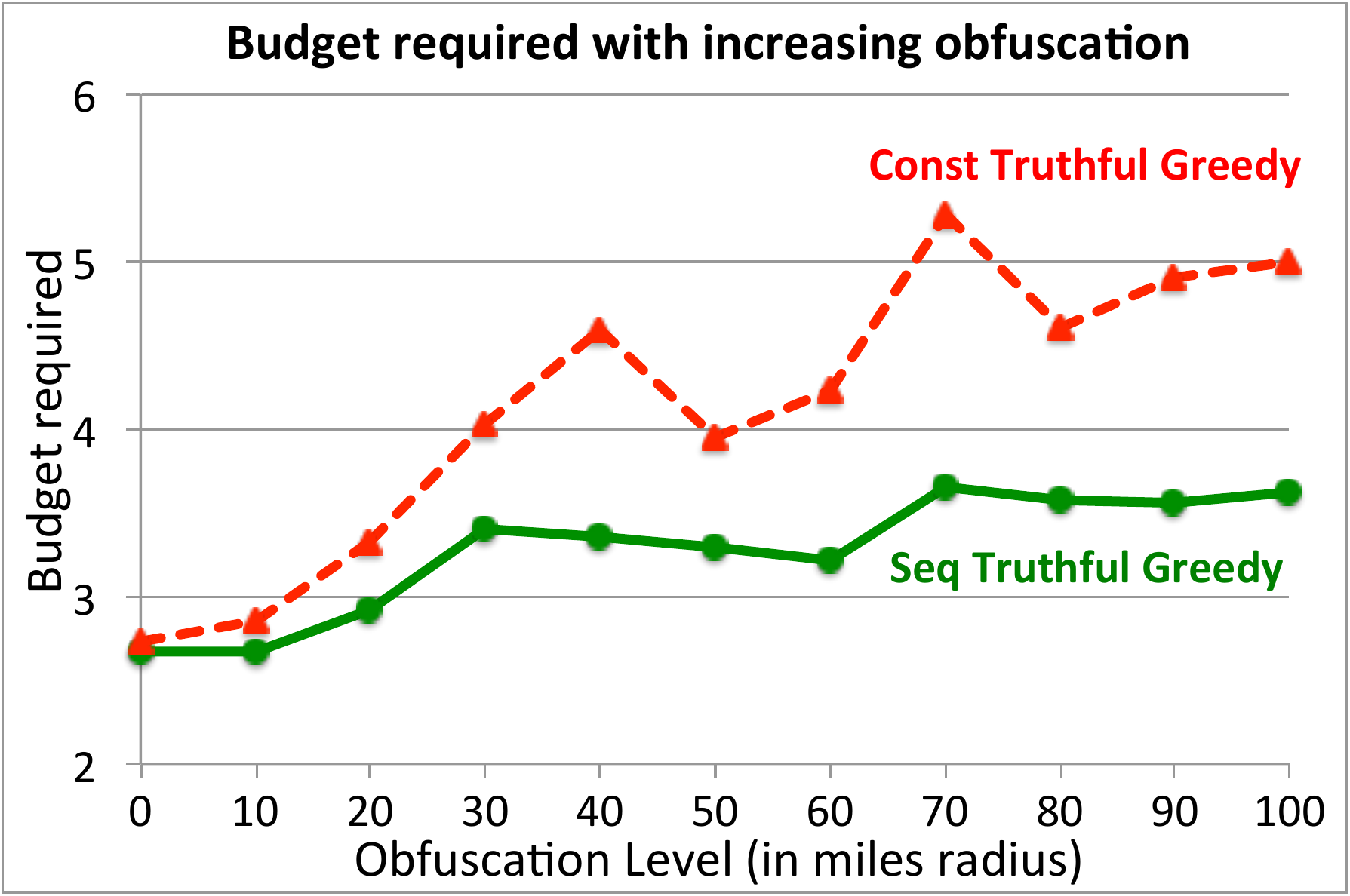}
    \label{fig:budget_vary-obfuscation}
   }
   \subfigure[\% Budget change: Varying obfuscation]{
     \includegraphics[width=0.32\textwidth]{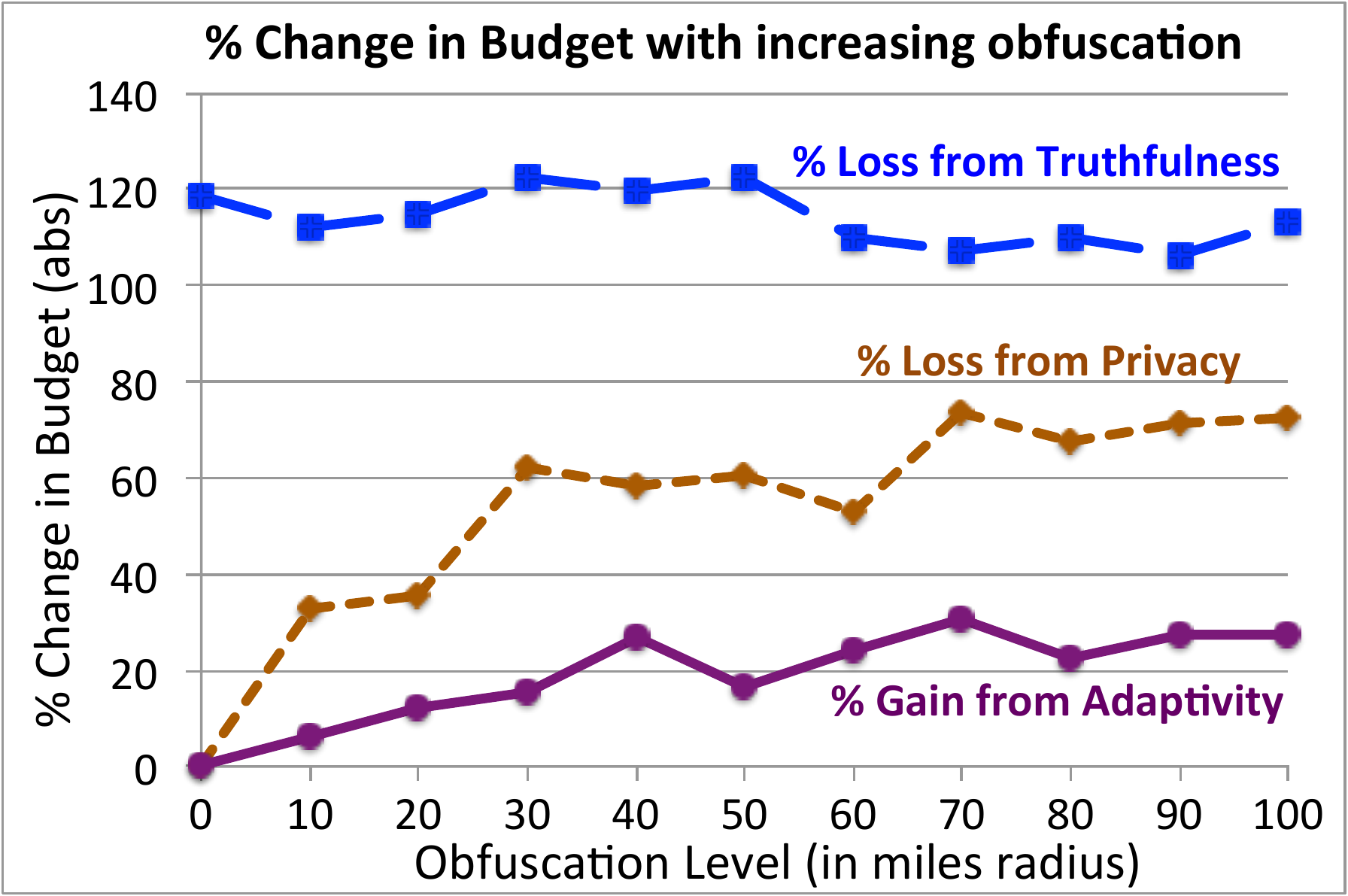}
     \label{fig:budget-change_vary-obfuscation}
   }
\caption{In ({\bf a}) and ({\bf d}), for a fixed obfuscation level of 100 miles radius, budget given and desired utility are varied. In ({\bf b}), ({\bf c}), ({\bf e}) and ({\bf f})) the obfuscation level is varied. ({\bf b}) and ({\bf c}) measure utility acquired for a given budget of 5\$ and show about  \textit{5\% adaptivity gain}. ({\bf e}) and ({\bf f}) measure the budget required (in \$) to achieve a utility of 120 and show up to \textit{30\% adaptivity gain}.}
\label{fig:results_all-metrics}
\end{figure*}

In community sensing applications with a large number of users and bounded maximal contribution from each user, $\alpha$ is close to 1, resulting in a utilization of almost the entire budget. Figure~\ref{fig:budget_vary-alpha} demonstrates the benefit of using tighter payment bounds (optimized $\alpha$), compared to a mechanism simply using $\alpha = 2$. Henceforth, in the results, we use the optimized $\alpha$ for all the truthful mechanisms.

{\bf Varying the given budget and specified utility.}
For a fixed obfuscation level of 100 miles radius, Figures~\ref{fig:utility_vary-budget} and \ref{fig:budget_vary-utility} show the effect of varying the given budget and desired utility respectively. Figure~\ref{fig:utility_vary-budget} illustrates the bounded approximation of our mechanism \stgreedy w.r.t. \sgreedy and up to 5\% improvement over \ctgreedy in terms of acquired utility. Figure~\ref{fig:budget_vary-utility} shows that the budget required to achieve a specified utility by our mechanism is larger w.r.t. \sgreedy and we achieve up to 20\% reduction in  required budget by using the adaptive mechanism.

{\bf Utility acquired at different obfuscation levels.}
In Figures~\ref{fig:utility_vary-obfuscation} and~\ref{fig:utility-change_vary-obfuscation}, the acquired utility is measured for a given budget of 5\$ by varying the obfuscation level. We can see that adaptivity helps acquire about 5\% higher utility and this adaptivity gain increases with higher obfuscation (more privacy). The loss from truthfulness is bounded (by 25\%), agreeing with our approximation guarantees. The loss from the lack of private information grows, but so also does the gain from adaptivity, which helps to reduce the loss we incur due to privacy protection.

{\bf Budget required at different obfuscation levels.}
In Figures~\ref{fig:budget_vary-obfuscation} and~\ref{fig:budget-change_vary-obfuscation}, the  required budget is computed for a desired utility value of 120 by varying the obfuscation level. We can see an increasing adaptivity gain, up to a total of 30\% reduction in required budget. As the privacy level increases, the adaptivity gain increases to help partially recover the incurred loss from privacy in terms of budget requirement.
%
\section{Conclusions and Future Work}\label{sec:discussion}
There is much potential in intelligent systems that incentivize and empower their users to consciously share certain private information. We presented a principled approach for negotiating access to such private information in community sensing. By using insights from mechanism design and adaptive submodular optimization, we designed the first adaptive, truthful and budget feasible mechanism guaranteed to recruit a near-optimal subset of participants. We demonstrated the feasibility and efficiency of our approach in a realistic case study. 
Privacy tradeoff is a personal choice and sensitive issue. In realistic deployments of the proposed approach, the choice of participation ultimately lies with the users. We believe that this integrated approach connecting privacy, utility and incentives provides an important step towards developing practical, yet theoretically well-founded techniques for community sensing.

There are some natural extensions for future work. Here, we considered a fairly simple utility function for the sensing phenomena. More complex objectives, e.g., reduction in predictive variance in a statistical model, can be readily incorporated.
Further, we would like to design an application (\emph{e.g.}, smartphone app) for deploying our approach in a real world sensing application. 
It would also be interesting to apply our mechanisms to other application domains that involve uncertainty, sequential decision-making and strategic interactions, \emph{e.g.}, viral marketing.




\vspace{3mm}
\noindent {\footnotesize \textbf{Acknowledgments.}
We would like to thank Yuxin Chen and G\'abor Bart\'ok for helpful discussions. This research was supported in part by SNSF grant 200021\_137971,  ERC StG 307036 and a Microsoft Research Faculty Fellowship.}
\clearpage
{
\fontsize{9.5pt}{10.5pt}
\selectfont
\bibliographystyle{aaai}
\bibliography{incentives-privacy}  

\begin{thebibliography}{}

\bibitem[\protect\citeauthoryear{Aberer \bgroup et al\mbox.\egroup
  }{2010}]{2010-iwgs_opensense}
Aberer, K.; Sathe, S.; Chakraborty, D.; Martinoli, A.; Barrenetxea, G.;
  Faltings, B.; and Thiele, L.
\newblock 2010.
\newblock Opensense: Open community driven sensing of environment.
\newblock {\em IWGS}.

\bibitem[\protect\citeauthoryear{Asadpour, Nazerzadeh, and
  Saberi}{2008}]{2008-wine-journal-version_stochastic-submodular}
Asadpour, A.; Nazerzadeh, H.; and Saberi, A.
\newblock 2008.
\newblock Maximizing stochastic monotone submodular functions.

\bibitem[\protect\citeauthoryear{Carrascal \bgroup et al\mbox.\egroup
  }{2013}]{2013-www_your-browsing-behavior-for-a-big-mac}
Carrascal, J.~P.; Riederer, C.; Erramilli, V.; Cherubini, M.; and de~Oliveira,
  R.
\newblock 2013.
\newblock Your browsing behavior for a big mac: economics of personal
  information online.
\newblock WWW '13,  189--200.

\bibitem[\protect\citeauthoryear{Chen, Gravin, and
  Lu}{2011}]{2011-soda_improved-budget-feasible}
Chen, N.; Gravin, N.; and Lu, P.
\newblock 2011.
\newblock On the approximability of budget feasible mechanisms.
\newblock In {\em SODA}.

\bibitem[\protect\citeauthoryear{Chon \bgroup et al\mbox.\egroup
  }{2012}]{2012-ubicomp_cps}
Chon, Y.; Lane, N.~D.; Li, F.; Cha, H.; and Zhao, F.
\newblock 2012.
\newblock Automatically characterizing places with opportunistic crowdsensing
  using smartphones.
\newblock In {\em Ubicomp}.

\bibitem[\protect\citeauthoryear{Chorppath and
  Alpcan}{2012}]{2012-pmc_trading-privacy-with-incentives}
Chorppath, A.~K., and Alpcan, T.
\newblock 2012.
\newblock Trading privacy with incentives in mobile commerce: A game theoretic
  approach.
\newblock {\em Pervasive and Mobile Computing}.

\bibitem[\protect\citeauthoryear{Clayton \bgroup et al\mbox.\egroup
  }{2012}]{2012-geophysics_krause_community-seismic-network}
Clayton, R.; Heaton, T.; Chandy, M.; Krause, A.; Kohler, M.; Bunn, J.; Olson,
  M.; Faulkner, M.; Cheng, M.; Strand, L.; Chandy, R.; Obenshain, D.; Liu, A.;
  Aivazis, M.; and Guy, R.
\newblock 2012.
\newblock Community seismic network.
\newblock {\em Annals of Geophysics} 54(6):738--747.

\bibitem[\protect\citeauthoryear{Dwork}{2006}]{2006-icalp_differential-privacy}
Dwork, C.
\newblock 2006.
\newblock Differential privacy.
\newblock In {\em ICALP}, volume 4052,  1--12.

\bibitem[\protect\citeauthoryear{Faltings, Jurca, and
  Li}{2012}]{2012-internet-of-things_truthful-measurements}
Faltings, B.; Jurca, R.; and Li, J.~J.
\newblock 2012.
\newblock Eliciting truthful measurements from a community of sensors.
\newblock {\em 3rd Int. Conference on Internet of Things}  51--18.

\bibitem[\protect\citeauthoryear{Feige}{1998}]{1998-_feige_threshold-of-ln-n}
Feige, U.
\newblock 1998.
\newblock A threshold of ln n for approximating set cover.
\newblock {\em Journal of the ACM} 45:314--318.

\bibitem[\protect\citeauthoryear{Golovin and
  Krause}{2011}]{2011-jair_krausea_adaptive-submodularity}
Golovin, D., and Krause, A.
\newblock 2011.
\newblock Adaptive submodularity: Theory and applications in active learning
  and stochastic optimization.
\newblock {\em Journal of Artificial Intelligence Research (JAIR)} 42:427--486.

\bibitem[\protect\citeauthoryear{Himmel \bgroup et al\mbox.\egroup
  }{2005}]{2005-patent_schedule-based-ad-on-phone}
Himmel, M.; Rodriguez, H.; Smith, N.; and Spinac, C.
\newblock 2005.
\newblock Method and system for schedule based advertising on a mobile phone.

\bibitem[\protect\citeauthoryear{Horvitz \bgroup et al\mbox.\egroup
  }{2005}]{2005-uai_horvitz_traffic}
Horvitz, E.; Apacible, J.; Sarin, R.; and Liao, L.
\newblock 2005.
\newblock Prediction, expectation, and surprise: Methods, designs, and study of
  a deployed traffic forecasting service.
\newblock In {\em UAI}.

\bibitem[\protect\citeauthoryear{Hui \bgroup et al\mbox.\egroup
  }{2011}]{2011-hci_targeted-ad-on-phone}
Hui, P.; Henderson, T.; Brown, I.; and Haddadi, H.
\newblock 2011.
\newblock Targeted advertising on the handset : privacy and security
  challenges.
\newblock In {\em Pervasive Advertising}, HCI'11.

\bibitem[\protect\citeauthoryear{Kansal \bgroup et al\mbox.\egroup
  }{2007}]{2007-ieee_senseweb}
Kansal, A.; Nath, S.; Liu, J.; and Zhao, F.
\newblock 2007.
\newblock Senseweb: An infrastructure for shared sensing.
\newblock {\em IEEE Multimedia} 14(4).

\bibitem[\protect\citeauthoryear{Krause and
  Guestrin}{2007}]{2007-aaai_krause_observation-selection}
Krause, A., and Guestrin, C.
\newblock 2007.
\newblock Near-optimal observation selection using submodular functions.
\newblock In {\em AAAI, Nectar track}.

\bibitem[\protect\citeauthoryear{Krause and
  Horvitz}{2008}]{2008-aaai_krause_privacy-personalization}
Krause, A., and Horvitz, E.
\newblock 2008.
\newblock A utility-theoretic approach to privacy and personalization.
\newblock In {\em AAAI}.

\bibitem[\protect\citeauthoryear{Krause \bgroup et al\mbox.\egroup
  }{2008}]{2008-ispn_krause_community-sensing}
Krause, A.; Horvitz, E.; Kansal, A.; and Zhao, F.
\newblock 2008.
\newblock Toward community sensing.
\newblock In {\em IPSN}.

\bibitem[\protect\citeauthoryear{Krumm and
  Horvitz}{2006}]{2006-ubicomp_predestination}
Krumm, J., and Horvitz, E.
\newblock 2006.
\newblock Predestination: Inferring destinations from partial trajectories.
\newblock In {\em Ubicomp},  243--260.

\bibitem[\protect\citeauthoryear{Krumm}{2007}]{2007-pervasive_inference-attack}
Krumm, J.
\newblock 2007.
\newblock Inference attacks on location tracks.
\newblock In {\em PERVASIVE},  127--143.

\bibitem[\protect\citeauthoryear{Li and
  Faltings}{2012}]{2012-comp-sust_incentive-schemes-communitysensing}
Li, J.~J., and Faltings, B.
\newblock 2012.
\newblock Incentive schemes for community sensing.
\newblock {\em The 3rd International Conference in Computational
  Sustainability}.

\bibitem[\protect\citeauthoryear{Lieb}{2007}]{2007-_modot}
Lieb, D.~A.
\newblock 2007.
\newblock Mo{DOT} tracking cell phone signals to monitor traffic speed,
  congestion.

\bibitem[\protect\citeauthoryear{Liu, Krishnamachari, and
  Annavaram}{2008}]{2008-MSWiM_game-theory-location-sharing}
Liu, H.; Krishnamachari, B.; and Annavaram, M.
\newblock 2008.
\newblock Game theoretic approach to location sharing with privacy in a
  community based mobile safety application.
\newblock In {\em MSWiM},  229--238.

\bibitem[\protect\citeauthoryear{Machanavajjhala \bgroup et al\mbox.\egroup
  }{2006}]{2006-icde_l-diversity}
Machanavajjhala, A.; Kifer, D.; Gehrke, J.; and Venkitasubramaniam, M.
\newblock 2006.
\newblock L-diversity: Privacy beyond k-anonymity.
\newblock In {\em ICDE}.

\bibitem[\protect\citeauthoryear{Mobile-Millennium}{2008}]{2008-online_mobile-millennium}
Mobile-Millennium.
\newblock 2008.
\newblock Mobile millennium traffic-monitoring system.
\newblock \url{http://traffic.berkeley.edu/}.

\bibitem[\protect\citeauthoryear{Myerson}{1981}]{1981-mor_myerson_optimal-auction-design}
Myerson, R.
\newblock 1981.
\newblock Optimal auction design.
\newblock {\em Mathematics of Operations Research} 6(1).

\bibitem[\protect\citeauthoryear{Nemhauser, Wolsey, and
  Fisher}{1978}]{1978-_nemhauser_submodular-max}
Nemhauser, G.; Wolsey, L.; and Fisher, M.
\newblock 1978.
\newblock An analysis of the approximations for maximizing submodular set
  functions.
\newblock {\em Math. Prog.} 14:265--294.

\bibitem[\protect\citeauthoryear{Olson, Grudin, and
  Horvitz}{2005}]{2005-chi_privacy-preferences}
Olson, J.; Grudin, J.; and Horvitz, E.
\newblock 2005.
\newblock A study of preferences for sharing and privacy.
\newblock In {\em CHI}.

\bibitem[\protect\citeauthoryear{Singer}{2010}]{2010-focs_singer_budget-feasible-mechanisms}
Singer, Y.
\newblock 2010.
\newblock Budget feasible mechanisms.
\newblock In {\em FOCS},  765--774.

\bibitem[\protect\citeauthoryear{Singer}{2012}]{2012-wsdm_singer_how-to-win-friends-and-influence-people}
Singer, Y.
\newblock 2012.
\newblock How to win friends and influence people, truthfully: Influence
  maximization mechanisms for social networks.
\newblock In {\em WSDM}.

\bibitem[\protect\citeauthoryear{Singla and
  Krause}{2013a}]{2013-arxiv_singla_incentives-privacy}
Singla, A., and Krause, A.
\newblock 2013a.
\newblock Incentives for privacy tradeoff in community sensing (extended
  version).
\newblock \url{http://arxiv.org/abs/1308.4013}.

\bibitem[\protect\citeauthoryear{Singla and
  Krause}{2013b}]{2013-www_singla_truthful-incentives}
Singla, A., and Krause, A.
\newblock 2013b.
\newblock Truthful incentives in crowdsourcing tasks using regret minimization
  mechanisms.
\newblock WWW '13,  1167--1178.

\bibitem[\protect\citeauthoryear{Sviridenko}{2004}]{2004-operations_sviridenko_budgeted-submodular-max}
Sviridenko, M.
\newblock 2004.
\newblock A note on maximizing a submodular set function subject to knapsack
  constraint.
\newblock {\em Operations Research Letters} v.(32):41--43.

\bibitem[\protect\citeauthoryear{Sweeney}{2002}]{2002-journal-ufks_k-anonymity}
Sweeney, L.
\newblock 2002.
\newblock k-anonymity: a model for protecting privacy.
\newblock {\em Int. Journal on Uncertainty, Fuzziness and Knowledge-based
  Systems} 10(5):557--570.

\bibitem[\protect\citeauthoryear{Wunnava \bgroup et al\mbox.\egroup
  }{2007}]{2007-techreport_travel-time-estimation}
Wunnava, S.; Yen, K.; Babij, T.; Zavaleta, R.; Romero, R.; and Archilla, C.
\newblock 2007.
\newblock Travel time estimation using cell phones {(TTECP)} for highways and
  roadways.
\newblock Technical report, Florida Department of Transportation.

\bibitem[\protect\citeauthoryear{Yoon, Noble, and
  Liu}{2007}]{2007-mobisys_street-traffic-estimation}
Yoon, J.; Noble, B.; and Liu, M.
\newblock 2007.
\newblock Surface street traffic estimation.
\newblock In {\em MobiSys},  220--232.

\bibitem[\protect\citeauthoryear{Zheng, Xie, and
  Ma}{2010}]{2010-ieee_msr_geolife}
Zheng, Y.; Xie, X.; and Ma, W.-Y.
\newblock 2010.
\newblock Geolife: A collaborative social networking service among user,
  location and trajectory.
\newblock {\em IEEE Data Engineering Bulletin}  32--40.

\end{thebibliography}
}
%
\clearpage
\appendix 
{\allowdisplaybreaks
\section{Proof of Theorem~\ref{theorem:truthful}}
Let $\icS$ denote the set of participants allocated by $\iPolicy_\iMechanism$ along with making observations $\ibobs_\icS$. We use $\ibobsSet_{\icW, \icS} = [\ibobs^1, \ibobs^2 \dots \ibobs^\ir \dots \ibobs^\iZ]$, where $\iZ = |\ibobsSet_{\icW, \icS}|$, to denote the set of possible realizations of $\ibObs_\icW = \ibobs_\icW \subseteq \icW \times \icO$ consistent with $\ibobs_\icS$.
In Lemma~\ref{lemma:truthful:deterministic}, we first prove the truthfulness of the payment $\ipaymentconst_{\is}(\ibobs^\ir)$ made for each of these possible realizations $\ibobs^\ir \in \ibobsSet_{\icW, \icS}$ (also denoted as $\ipaymentconstr_{\is}$).
To prove Lemma~\ref{lemma:truthful:deterministic}, we first show  allocation rule is monotone (Lemma~\ref{lemma:truthful:deterministic:monotone}) and allocated users are paid threshold payments (Lemma~\ref{lemma:truthful:deterministic:threshold}). 
\begin{lemma} \label{lemma:truthful:deterministic:monotone}
For a given $\ibobs_\icW$, allocation policy of the mechanism is monotone i.e $\forall \ii \in [\iin] \text{ and for every } \ibid_{-\ii}, \text{ if } \ibidalt_{\ii}  \leq \ibid_\ii \text{ then } \ii \in \ipolicy(\ibid_\ii, \ibid_{-\ii}) \text{ implies } \ii \in \ipolicy(\ibidalt_{\ii}, \ibid_{-\ii})$
\end{lemma}
\begin{proof}[\bf{Proof}]
\looseness-1 The monotonicity of the greedy scheme is easy to see:  By lowering her bid, any allocated participant  would only increase their marginal gain per unit cost and thus jump ahead in the sorting order considered by the allocation policy.
\end{proof}
\begin{lemma} \label{lemma:truthful:deterministic:threshold}
Payment $\ipaymentconst_{\is}$ for a given $\ibobs_\icW$ is a threshold payment, {\em i.e.}, payment to each winning bidder is $\operatorname*{inf} \{\ibidalt_\ii : \ii \notin \ipolicy(\ibidalt_\ii, \ibid_{-\ii})\}$
\end{lemma}
\begin{proof}[\bf{Proof}]
The threshold payment for participant $\is = \ii$ is given by $\ipaymentconst_{\ii} = \max_{j \in [\ikalt + 1]}(\ipaymentconst_{\ii(\iij)})$ where $\ipaymentconst_{\ii(\iij)} = \min(\ibid_{\ii(\iij)}, \irho_{\ii(\iij)})$ as the bid that $\ii$ can declare to replace $\iij$ in $\icSalt$. We have $\ibid_{\ii(\iij)} = \frac{\imarginal_{\ii(\iij)} \cdot \ibid_\iij}{\imarginalalt_\iij}$ and $\irho_{\ii(\iij)} = \frac{\icB}{\ialpha} \cdot \frac{\imarginal_{\ii(\iij)}}{\sum_{\isalt \in [\iij-1]} \imarginalalt_{\isalt} + \imarginal_{\ii(\iij)}}$. Let us consider $\ir$ to be the index for which $\ipaymentconst_{\ii} = \min(\ibid_{\ii(\ir)}, \irho_{\ii(\ir)})$. Declaring a bid of $\min(\ibid_{\ii(\ir)}, \irho_{\ii(\ir)})$ ensures that $\is$ would definitely get allocated at position $\ir$ in the alternate run of the policy. Let us consider the following four cases:

{\bf Case 1: $\ibid_{\ii(\ir)} \leq \irho_{\ii(\ir)} \ \& \ \ibid_{\ii(\ir)} = \max_\iij{\ibid_{\ii(\iij)}} $}\\
Reporting a bid higher than $\ibid_{\ii(\ir)}$ places the $\ii$ after the unalocated user $\ikalt + 1$ in the alternate run of the mechanism, thereby $\ii$ would not be allocated.

{\bf Case 2: $\ibid_{\ii(\ir)} \leq \irho_{\ii(\ir)} \ \& \ \ibid_{\ii(\ir)} < \max_\iij{\ibid_{\ii(\iij)}} $}\\
Consider some $\iij$ for which $\ibid_{\ii(\ir)} < \ibid_{\ii(\iij)}$. Because of the maximal condition for $\ir$, it must be the case that $\irho_{\ii(\iij)} \leq \ibid_{\ii(\ir)} \leq \ibid_{\ii(\iij)}$. Thus, declaring a bid higher than $\ibid_{\ii(\ir)}$ would violate the proportional share allocation condition and hence $\ii$ would not be allocated. For some other $\iij$ for which $\ibid_{\ii(\ir)} \geq \ibid_{\ii(\iij)}$, declaring a bid higher than $\ibid_{\ii(\ir)}$ would put $\ii$ after $\iij$ and hence $\ii$ would not be allocated at considered position $\iij$.

{\bf Case 3: $\irho_{\ii(\ir)} \leq \ibid_{\ii(\ir)} \ \& \ \irho_{\ii(\ir)} = \max_\iij{\irho_{\ii(\iij)}} $}\\
Reporting a bid higher than $\irho_{\ii(\ir)}$ violates the proportional share allocation condition at each of the indices in $j \in [\ikalt + 1]$, hence $\ii$ would not be allocated.

{\bf Case 4: $\irho_{\ii(\ir)} \leq \ibid_{\ii(\ir)} \ \& \ \irho_{\ii(\ir)} < \max_\iij{\irho_{\ii(\iij)}} $}\\
Consider some $\iij$ for which $\irho_{\ii(\ir)} < \irho_{\ii(\iij)}$. Because of the maximal condition for $\ir$, it must be the case that $\ibid_{\ii(\iij)} \leq \irho_{\ii(\ir)} \leq \irho_{\ii(\iij)}$. Thus, declaring a bid higher than $\irho_{\ii(\ir)}$ would put $\ii$ after $\iij$  and hence $\ii$ would not be allocated. For any other $\iij$ for which $\irho_{\ii(\ir)} \geq \irho_{\ii(\iij)}$, declaring a bid higher than $\ibid_{\ii(\ir)}$ would violate the proportional share allocation condition and hence $\ii$ would not be allocated at considered position $\iij$.

The anaylysis of above four cases completes the proof.
\end{proof}
\begin{lemma} \label{lemma:truthful:deterministic}
Payment $\ipaymentconst_{\is}$ for a given $\ibobs_\icW$ is truthful.
\end{lemma}
\begin{proof}[\bf{Proof}]
To prove this, we use the well-known characterization of \citet{1981-mor_myerson_optimal-auction-design}. For the case of deterministic settings in single parameter domains, a mechanism is truthful if the allocation rule is monotone and the allocated agents are paid threshold payments. 
\end{proof}
\begin{proof}[\bf{Proof of Theorem \ref{theorem:truthful}}]
The final payment made to participant $\is$ is given by $\ipayment_\is = \sum_{\ibobs^\ir \in \ibobsSet_{\icW, \icS}} \iP(\ibObs_\icW = \ibobs^\ir|\ibobs_{\icS}) \cdot \ipaymentconstr_{\is}$. From Lemma~\ref{lemma:truthful:deterministic}, each of the payments $\ipaymentconstr_{\is}$ are truthful, {\em i.e.}, the profit of a user cannot be increased by deviating from their true cost. Taking a linear combination of these payments ensures truthful payment as well.
\end{proof}
\section{Proof of Theorem~\ref{theorem:rational}}
In Lemma~\ref{lemma:rational:deterministic}, we first prove the individual rationality of the payment $\ipaymentconst_{\is}(\ibobs^\ir)$ made for each of these possible realizations $\ibobs^\ir \in \ibobsSet_{\icW, \icS}$ (also denoted as $\ipaymentconstr_{\is}$).
\begin{lemma} \label{lemma:rational:deterministic}
Payment $\ipaymentconst_{\is}$ for a given $\ibobs_\icW$ is individually rational i.e. $\ipaymentconst_\is \geq \ibid_\is$
\end{lemma}
\begin{proof}[\bf{Proof}]
Consider the bid that $\ii$ can declare to be allocated at position $\iij = \ii$ (i.e. back at its original position) in the alternate run of the mechanism. $\ipaymentconst_{\ii(\ii)} = \min(\ibid_{\ii(\ii)}, \irho_{\ii(\ii)})$. We will show that $\ibid_\ii \leq \ipaymentconst_{\ii(\ii)}$.

{\bf Showing $\ibid_{\ii(\ii)} \geq \ibid_\ii$}\\
\begin{align*}
\ibid_{\ii(\ii)} &= \frac{\imarginal_{\ii(\ii)} \cdot \ibid_\iij}{\imarginalalt_\iij}
= \frac{\imarginal_{\ii} \cdot \ibid_\iij}{\imarginal_\iij} \tag{1} \\
&\geq \frac{\imarginal_{\ii} \cdot \ibid_\ii}{\imarginal_\ii} 
= \ibid_\ii \tag{2}
\end{align*}
In step 1, the second equality holds from the fact that the first $\ii-1$ allocated elements in both runs of the policies are the same and hence $\imarginal_{\ii(\ii)} = \imarginal_{\ii}$ and $\imarginalalt_\iij = \imarginal_\iij$. In step 2, the first inequality holds from the fact that $\frac{\ibid_\iij}{\imarginal_\iij} \geq \frac{\ibid_\ii}{\imarginal_\ii}$ since $\ii$ was allocated in the original run of the policy after $\ii-1$, instead of user $\iij$.

{\bf Showing $\irho_{\ii(\ii)} \geq \ibid_\ii$}\\
\begin{align*}
\irho_{\ii(\ii)} &= \frac{\icB}{\ialpha} \cdot \frac{\imarginal_{\ii(\ii)}}{\sum_{\isalt \in [\ii-1]} \imarginalalt_{\isalt} + \imarginal_{\ii(\ii)}} \\
&= \frac{\icB}{\ialpha} \cdot \frac{\imarginal_{\ii}}{\sum_{\is \in [\ii-1]} \imarginal_{\is} + \imarginal_{\ii}} \geq \ibid_\ii \tag{3}
\end{align*}
In step 3, the first equality holds from the fact that the first $\ii-1$ allocated elements in both the runs of the policies are same. The second inequality follows from the proportional share creteria used to decide the allocation of $\ii$ after $\ii-1$ users were allocated already.

Now, we have $\ibid_\ii \leq \ipaymentconst_{\ii(\ii)} \leq \max_{j \in [\ikalt + 1]}(\ipaymentconst_{\ii(\iij)}) = \ipaymentconst_{\ii}$
\end{proof}
\begin{proof}[\bf{Proof of Theorem \ref{theorem:rational}}]
The final payment made to participant $\is$ is given by $\ipayment_\is = \sum_{\ibobs^\ir \in \ibobsSet_{\icW, \icS}} \iP(\ibObs_\icW = \ibobs^\ir|\ibobs_{\icS}) \cdot \ipaymentconstr_{\is}$. From Lemma~\ref{lemma:rational:deterministic}, each of the payment $\ipaymentconstr_{\is} \geq \ibid_\is$. Taking a linear combination of these payments ensures individual rationality in expectation as well.
\end{proof}
\section{Proof of Theorem~\ref{theorem:budgetfeasible}}
The theorem rests on the following Lemma~\ref{lemma:budgetfeasible:boundpayment} which upper bounds the payments made to each participant by $\alpha_{\geq 1}$ times their marginal contribution to the total utility of the final set of participants. 

\begin{lemma} \label{lemma:budgetfeasible:boundpayment}
When full budget $\icB$ is used by mechanism, the maximum raise in bid $\ibidalt_\is$ that a participant $\is$ can make, keeping the bids of others same,  to still get selected by mechanism is upper bounded by $\ialpha \cdot \frac{\imarginal_\is}{\sum_{\is \in \icS} \imarginal_{\is}} \cdot \icB$ where $\ialpha \leq 2$.
\end{lemma}
\begin{proof}[\bf{Proof}]
Consider any random realization $\ibObs_\icW = \ibobs_\icW$. Let $\icS$ be the set of participants selected by policy alongwith making observations $\ibobs_{\icS}$. 
Let us renumber the users in which they were allocated by mechanism $\icS = \{1,2,\dots,\ii-1,\ii (=\is),\dots,\ik\}$ and let's analyze the upper bound on the threshold payment for participant $\is = \ii$. Irrespective of the payment scheme used, we consider how much raised bid participant $\ii$ ($\ibidalt_\ii$ raised from $\ibid_\ii$) can declare to still selected by the mechanism, keeping the bids of other users ($\ibid_{-\ii}$) same. We use $\iBid = (\ibid_\ii, \ibid_{-\ii})$ to denote original bids and $\iBidalt = (\ibidalt_\ii, \ibid_{-\ii})$ to denote modified bids. Consider running policy on alternate bids $\iBidalt$ and let $\icSalt = \{1,2,\dots,\iij-1, \iij (=\is),\dots,\ikalt\}$ be the allocated set (users again renumbered based on order of allocation). For distintion, we use $\imarginal$ and $\imarginalalt$ to denote the marginal contributions of the users in the above two different runs of the policy. Let $\icTalt$ denote the subset of participants from $\icSalt$ which were allocated just before $\is$ was allocated at position $\iij$. Let us consider following two cases:

{\bf Case 1: $\icS \setminus \icTalt = \emptyset$}.\\
This condition also implies that $\icTalt \cup \{\is\} = \icTalt \cup \icS$. Let $\imarginalalt(\is | \ibobs_\icTalt)$ denote marginal contribution of $\is$ when added by policy after $\icTalt$. We have 
\begin{align*}
\ibidalt_\ii &\leq \icB \cdot \frac{\imarginalalt(\is | \ibobs_\icTalt)}{\iobjg(\ibobs_\icTalt \cup \{\is, \iobs_\is\})}
             = \icB \cdot \frac{\imarginalalt(\is | \ibobs_\icTalt)}{\iobjg(\ibobs_\icTalt \cup \ibobs_\icS)} \tag{1}\\
             &\leq \icB \cdot \frac{\imarginalalt(\is | \ibobs_\icTalt)}{\iobjg(\ibobs_\icS)} 
			 \leq \icB \cdot \frac{\imarginal_\is}{\iobjg(\ibobs_\icS)} \tag{2}
\end{align*}
Setting $\ibidalt_\ii = \alpha \cdot \icB \cdot \frac{\imarginal_\is}{\iobjg(\ibobs_\icS)}$, we get
\begin{align*}
\ialpha &= 1 \tag{3}
\end{align*}
First inequality in step 1 follows from the propotional share allocation creteria and second equality follows from the fact that  $\icTalt \cup \{\is\} = \icTalt \cup \icS$. In step 2,  first inquality follows from monotonicity of function $\iobjg$ and second inequality follows from the fact that increasing the bid by $\is$ can only pushes her position lower in the allocation, decreasing the marginal contribution. Note that here $\imarginal_\is$ is used to denote the marginal contribution of $\is$ when it was allocated at position $\ii$ in the original run of the policy. Finally, in step 3, the inequality holds for ${\ialpha = 1}$.  

{\bf Case 2: $\icS \setminus \icTalt = \icR$}\\
We have
\begin{align*}
\ibidalt_\ii &\leq \icB \cdot \frac{\imarginalalt(\is | \ibobs_\icTalt)}{\iobjg(\ibobs_\icTalt \cup \{\is, \iobs_\is\})}
              \leq \icB \cdot \frac{\imarginal_\is}{\iobjg(\ibobs_\icTalt \cup \{\is, \iobs_\is\})} \tag{4}
\end{align*}
Setting $\ibidalt_\ii = \alpha \cdot \icB \cdot \frac{\imarginal_\is}{\iobjg(\ibobs_\icS)}$, we get
\begin{align*}
\frac{\iobjg(\ibobs_\icTalt \cup \{\is, \iobs_\is\})}{\iobjg(\ibobs_\icS)} \leq \frac{1}{\ialpha} \tag{5}
\end{align*}
Now, consider adding some user on top of $\ibobs_\icTalt \cup \{\is, \iobs_\is\}$. For some $\ir_0 \in \icR$, it must hold that marginal value by unit cost of adding $\ir_0$ is higher than that of addding whole  $\icR$. We have,
\begin{align*}
&\frac{\iobjg(\ibobs_\icR \cup \ibobs_\icTalt \cup \{\is, \iobs_\is\}) - \iobjg(\ibobs_\icTalt \cup \{\is, \iobs_\is\})}{\iBidalt(\icR)} \\
&\leq \frac{\imarginalalt(\ir_0 | \ibobs_\icTalt \cup \{\is, \iobs_\is\})}{\ibidalt_{\ir_0}}\\
&\leq \frac{\imarginalalt(\ir_0 | \ibobs_\icTalt)}{\ibidalt_{\ir_0}}
\leq \frac{\imarginalalt(\is | \ibobs_\icTalt)}{\ibidalt_{\ii}} \tag{6}\\
&\leq \frac{\imarginal_\is}{\ibidalt_{\ii}} 
= \frac{\iobjg(\ibobs_\icS)}{\ialpha \cdot \icB} \tag{7}
\end{align*}
In step 6, first inequality holds from submodularity of $\iobjg$ and second holds from that fact that $\is$ was choosen to be added on set $\icTalt$ compared to $\ir_0$ at position $j$ by the alternate run of the mechansim. In step 7, first inequality follows from the fact that increasing the bid by $\is$ can only pushes her position lower in the allocation, decreasing the marginal contribution. The second inequality holds by setting $\ibidalt_\ii = \alpha \cdot \icB \cdot \frac{\imarginal_\is}{\iobjg(\ibobs_\icS)}$.

Now, using the fact that $\iBidalt(\icR) \leq \icB$, and $\iobjg(\ibobs_\icS) \leq \iobjg(\ibobs_\icS \cup \ibobs_\icTalt) = \iobjg(\ibobs_\icR \cup \ibobs_\icTalt \cup \{\is, \iobs_\is\})$, we have
\begin{align*}
&\frac{\iobjg(\ibobs_\icS ) - \iobjg(\ibobs_\icTalt \cup \{\is, \iobs_\is\})}{\icB} \leq \frac{\iobjg(\ibobs_\icS)}{\ialpha \cdot \icB} \tag{8}\\
&\frac{\iobjg(\ibobs_\icTalt \cup \{\is, \iobs_\is\})}{\iobjg(\ibobs_\icS)} \geq (1 - \frac{1}{\ialpha}) \tag{9}
\end{align*}
Combining step 5 and step 9, we get an upper bound on $\ialpha = 2$.
\end{proof}
\begin{proof}[\bf{Proof of Theorem \ref{theorem:budgetfeasible}}]
Consider running the mechanism with reduced budget of $\frac{\icB}{2}$ (i.e. seting parameter $\alpha = 2$ in the mechanism). Let a set $\icS$ allocated by mechanism and $\iPayment_\icS$ be the payments made to participants. By summing over these payments, we get :
\begin{align*}
\sum_{\is \in \icS} \ipayment_\is \leq 
\sum_{\is \in \icS} \ialpha \cdot \frac{\imarginal_\is}{\sum_{\is \in \icS} \imarginal_{\is}} \cdot \frac{\icB}{2} 
\leq \icB.
\end{align*} 
The inequality here holds from Lemma~\ref{lemma:budgetfeasible:boundpayment} which bounds the maximum threshold payment for a participant ${\is}$ by $\ialpha \leq 2$.
\end{proof}
\section{Proof of Theorem~\ref{theorem:approximation}}
Proof of Theorem~\ref{theorem:approximation} rests on proving following two lemmas. In Lemma~\ref{lemma:approximation:greedybound}, we first prove an upper bound on the utility of optimal sequential (untruthful) mechanism \sopt as $\sfrac{e}{(e - 1)}$ times the  utility on sequential greedy mechanism \sgreedy, with an extra additive factor of $\ifmax$. Then, in Lemma~\ref{lemma:approximation:truthgreedybound}, we show that, because of diminishing returns property of the utility functions, the stopping criteria used by the mechanism based on proportional share and using only $\ialpha$ proportion of the budget still allows the allocation of sufficiently many participants to achieve a competitive amount of utility for the application. Additionally, we use the fact that in our settings, the utility contribution of each participant is small compared to the overall utility achieved by the mechanism.

We use $\isoptpol$, $\isgreedypol$, $\istgreedypol$  to denote the allocation policies of mechanisms \sopt, \sgreedy and \stgreedy. Also, we use $\iobjgavg(\iPolicy)$ to denote the average expected utility obtained by running the allocation policy $\iPolicy$. We use the terms mechanism and policy interchangeably whenever clear from the context.
\begin{lemma} \label{lemma:approximation:greedybound}
Expected utility of optimal sequential policy $\sopt$ is bounded by the utility of sequential greedy policy \sgreedy as $\iobjgavg(\isoptpol) \leq \sfrac{e}{(e - 1)} \big[\iobjgavg(\isgreedypol) + \ifmax \big]$.
\end{lemma}
\begin{proof}[\bf{Proof}]
Let $\isgreedypol$ executes for $\il$ steps allocating a set $\icS_\il$. Let us renumber the users in order of which they were considered during execution of $\isgreedypol$ and denote $\icS_{\il + 1} = \{1,2,\dots,\ii-1,\ii,\dots,\il,\il + 1\}$ where $\il\ + 1$ is the first unallocated user because of budget constraint. Consider the step when participant $\ii$ is added by the policy on top of $\icS_{\ii - 1}$. We consider the expected marginal utility of executing the whole $\isoptpol$ after step $\ii - 1$, conditioned on observations $\ibobs_{\icS_{\ii-1}}$. 
Let $\ibobs_\icT$ be the final set of participants alongwith observations obtained by executing $\isoptpol$ after $\icS_{\ii - 1}$, where $\ibobs_{\icS_{\ii-1}} \subseteq \ibobs_\icT$. Let $\ir \in \icT \setminus \icS_{\ii -1}$. Given the submodularity of $\iobjg$, it must hold that:
\begin{align*}
&\frac{\iobjgavg(\ibobs_\icT \cup \ibobs_{\icS_{\ii - 1}}) - \iobjgavg(\ibobs_{\icS_{\ii - 1}})}{\iBid(\icT) - \iBid(\icS_{\ii - 1})} \leq \frac{\imarginal_\ir}{\ibid_\ir} \leq \frac{\imarginal_\ii}{\ibid_\ii}\tag{1}\\
&\frac{\iobjgavg(\isoptpol) - \iobjgavg(\ibobs_{\icS_{\ii - 1}})}{\icB} \leq \frac{\iobjgavg(\ibobs_{\icS_\ii}) - \iobjgavg(\ibobs_{\icS_{\ii - 1}})}{\ibid_\ii} \tag{2}\\
&\iobjgavg(\ibobs_{\icS_\ii}) \geq \frac{\ibid_\ii}{\icB} \cdot \iobjgavg(\isoptpol) + \big(1 - \frac{\ibid_\ii}{\icB} \big) \cdot \iobjgavg(\ibobs_{\icS_{\ii-1}}) \tag{3}
\end{align*}
Step 1 uses the fact that $\ii$ was choosen over $\ir$ by greediy policy. Step 2 uses the definition of $\imarginal_\ii = \iobjgavg(\ibobs_{\icS_\ii}) - \iobjgavg(\ibobs_{\icS_{\ii - 1}})$ and $\iobjgavg(\isoptpol) \leq \iobjgavg(\ibobs_\icT)$. By recursively applying step 3 results for $\il + 1$ steps, we get:
\begin{align*}
\iobjgavg(\ibobs_{\icS_{\il + 1}}) &\geq \Big[1 - \prod_{i \in [1 \dots \il+1]}\big(1 - \frac{\ibid_\ii}{\icB} \big) \Big] \cdot \iobjgavg(\isoptpol) \\
& \geq \Big[1 - \big(1 - \frac{\iBid(\icS_{\il+1})}{\icB} \cdot \frac{1}{\il+1} \big)^{\il + 1} \Big] \cdot \iobjgavg(\isoptpol) \tag{4} \\
& \geq \Big[1 - \big(1 - \frac{1}{\il+1} \big)^{\il + 1} \Big] \cdot \iobjgavg(\isoptpol) \tag{5} \\
& \geq (1 - \frac{1}{e}) \cdot \iobjgavg(\isoptpol) \tag{6}
\end{align*}
Step 4 uses the fact that minimum of the product for $\iin$ variables $\Big[1 - \prod_{i \in [1\dots \iin]}\big(1 - \frac{\ix_\ii}{\iX} \big) \Big]$ is achived when all the variables take value as $\ix_\ii = \frac{\iX}{\iin}$ (where $\iX = \sum_{i \in [1 \dots \iin]} \ix_\ii$). Step 5 uses the fact that $\iBid(\icS_{\il+1}) > \icB$ and Step 6 uses the limiting value of the equation.
\begin{align*}
\iobjgavg(\isoptpol) \leq (\frac{e}{e-1}) \cdot \big( \iobjgavg(\ibobs_{\icS_\il}) + \imarginal_{\il + 1} \big) \tag{7} \\
                     \leq (\frac{e}{e-1}) \cdot \big( \iobjgavg(\isgreedypol) + \ifmax \big) \tag{8}
\end{align*}
In step 7, we used the fact that $\iobjgavg(\ibobs_{\icS_{\il + 1}}) = \iobjgavg(\ibobs_{\icS_\il}) + \imarginal_{\il + 1}$. In step 8, we used the fact that $\imarginal_{\il+1} \leq \ifmax$ and $\iobjgavg(\isgreedypol) = \iobjgavg(\ibobs_{\icS_\il})$.
\end{proof}
\begin{lemma} \label{lemma:approximation:truthgreedybound}
Expected utility of sequential greedy policy \sgreedy is bounded by the utility of truthful greedy policy \stgreedy as $\iobjgavg(\isgreedypol) \leq (1 + \ialpha) \iobjgavg(\istgreedypol) + \ialpha \ifmax$.
\end{lemma}
\begin{proof}[\bf{Proof}]
Let $\isgreedypol$ executes for $\il$ steps allocating a set $\icS_\il$ and $\istgreedypol$ terminates after $\ik \leq \il$ steps because of additional stopping creteria allocating a set $\icS_\ik \subseteq \icS_\il$. Let us renumber the users in order of which they were considered during execution of $\isgreedypol$ and denote $\icS_\il = \{1,2,\dots,\ik,\ik + 1,\dots,\il\}$. Since $\ik + 1$ was not allocated by the $\istgreedypol$, we have: $\ibid_{\ik+1} > \frac{\icB}{\ialpha} \cdot \frac{\imarginal_{\ik+1}}{(\sum_{\ii \in \icS_\ik} \imarginal_{\ii} + \imarginal_{\ik+1}})$. Also, because of decreasing marginal utility by cost ratio of the users considered by the policy, we get:
\begin{align*}
&\frac{\ibid_\il}{\imarginal_\il} \geq \dots \geq \frac{\ibid_\iij}{\imarginal_\iij} \geq \dots \geq \frac{\ibid_{\ik+1}}{\imarginal_{\ik+1}} > \frac{\icB}{\ialpha} \cdot \frac{1}{(\sum_{\ii \in \icS_\ik} \imarginal_{\ii} + \imarginal_{\ik+1})} \\
&\implies \forall \iij \in [\ik+1 \dots \il], \ibid_\iij > \frac{\icB}{\ialpha} \cdot \frac{\imarginal_\iij}{(\sum_{\ii \in \icS_\ik} \imarginal_{\ii} + \imarginal_{\ik+1})} \\
&\icB \geq \sum_{\iij \in [\ik+1 \dots \il]}{\ibid_\iij} > \frac{\icB}{\ialpha} \cdot \frac{\sum_{\iij \in [\ik+1 \dots \il]}{\imarginal_\iij}}{(\sum_{\ii \in \icS_\ik} \imarginal_{\ii} + \imarginal_{\ik+1})} \\
&\ialpha \cdot (\iobjgavg(\istgreedypol) + \imarginal_{\ik+1}) \geq (\iobjgavg(\isgreedypol) - \iobjgavg(\istgreedypol)) \tag{9} \\
&\iobjgavg(\isgreedypol) \leq (1 + \ialpha) \iobjgavg(\istgreedypol) + \ialpha \ifmax \tag{10}
\end{align*} 
In step 9, we used the fact that $\iobjgavg(\isgreedypol) = \sum_{\ii \in \icS_\il} \imarginal_{\ii}$ and $\iobjgavg(\istgreedypol) = \sum_{\ii \in \icS_\ik} \imarginal_{\ii}$. In step 10, we used the fact that $\imarginal_{\ik+1} \leq \ifmax$.
\end{proof}
\begin{proof}[\bf{Proof of Theorem \ref{theorem:approximation}}]
Combining the results of above two lemmas, we get: 
\begin{align*}
\iobjgavg(\isoptpol) &\leq \frac{(1 + \ialpha) \cdot e}{e - 1} \big[ \iobjgavg(\istgreedypol) + \ifmax \big] \\
&= \frac{(1 + \ialpha) \cdot e}{e - 1} \Big(1 + \frac{\ifmax}{\iobjgavg(\istgreedypol)} \Big) \cdot \iobjgavg(\istgreedypol)
\end{align*} 
Now, we set $\ialpha = 2$. Also, using the fact that $\frac{\ifmax}{\iobjgavg(\istgreedypol)} \ll 1$ (i.e. each user can only contribute to a maximal of $\ifmax$ utility to the application which, for a large-scale application, is very small compared to utility achieved by mechanism under given budget), we get an approximation factor of $\sfrac{1}{4.75}$ $(=0.22)$.
\end{proof}

}

\end{document}